\documentclass[aoas,nameyear,dvips]{arximspdf}
\usepackage{graphics}
\doi{10.1214/09-AOAS285}
\volume{4}
\issue{1}
\pubyear{2010}
\firstpage{266}
\lastpage{298}

\makeatletter
\let\epsilon\varepsilon
\newcommand{\citeasnoun}[1]{\citet{#1}}
\makeatother

\begin{document}
\begin{frontmatter}

\title{BART: Bayesian additive regression trees\protect\thanksref{TIT1,TIT2}}
\runtitle{BART}
\thankstext{TIT1}{Preliminary versions of this paper were disseminated as June 2006 and June 2008 technical reports.}
\thankstext{TIT2}{Supported by NSF Grant DMS-06-05102, NSERC and the Isaac Newton Institute for Mathematical Sciences.}

\begin{aug}
\author[A]{\fnms{Hugh A.} \snm{Chipman}\ead[label=e1]{hugh.chipman@acadiau.ca}},
\author[B]{\fnms{Edward I.} \snm{George}\ead[label=e2]{edgeorge@wharton.upenn.edu}\corref{}}
\and
\author[C]{\fnms{Robert E.} \snm{McCulloch}\ead[label=e3]{robert.mcculloch1@gmail.com}}

\runauthor{H. A. Chipman, E. I. George and R. E. M{\normalfont c}Culloch}
\affiliation{Acadia University, University of Pennsylvania and\\ University of Texas at Austin}
\address[A]{H. A. Chipman\\
Department of Mathematics and Statistics\\
Acadia University\\
Wolfville, Nova Scotia, B4P 2R6\\
Canada\\
\printead{e1}} 
\address[B]{E. I. George\\
Department of Statistics\\
The Wharton School\\
University of Pennsylvania\\
3730 Walnut St, 400 JMHH\\
Philadelphia, Pennsylvania 19104-6304\\
USA\\
\printead{e2}}
\address[C]{R. E. McCulloch\\
IROM Department\\
University of Texas at Austin\\
1 University Station, B6500\\
Austin, Texas 78712-1175\\
USA\\
\printead{e3}}
\end{aug}

\received{\smonth{6} \syear{2008}}
\revised{\smonth{7} \syear{2009}}

\begin{abstract}
We develop a Bayesian ``sum-of-trees'' model where each
tree is constrained by a regularization prior to be a weak learner,
and fitting and inference are accomplished via an iterative Bayesian
backfitting MCMC algorithm that generates samples from a posterior.
Effectively, BART is a nonparametric Bayesian regression approach
which uses dimensionally adaptive random basis elements.  Motivated by ensemble methods
in general, and boosting algorithms in particular,  BART is defined by a
statistical model: a prior and a likelihood.  This approach enables full posterior inference
including point and interval estimates of the unknown regression function as
well as the marginal effects of potential predictors.  By keeping track of predictor
inclusion frequencies, BART can also be used for model-free variable selection.
BART's many features are illustrated with a bake-off against competing methods
on 42 different data sets, with a simulation experiment and on a
drug discovery classification problem.
\end{abstract}


\end{frontmatter}

\section{Introduction}\label{sec:intro}

We consider the fundamental problem of making inference about an
unknown function $f$ that predicts an output $Y$ using a $p$
dimensional vector of inputs $x = (x_1,\dots,x_p)$ when
\begin{equation}
Y=f(x) + \epsilon, \qquad  \epsilon \sim N(0,\sigma^2).
\label{basemodel}
\end{equation}
To do this, we consider modeling or at least approximating $f(x)
= E(Y | x)$, the mean of $Y$ given $x$, by a sum of $m$
regression trees $f(x) \approx h(x) \equiv \sum_{j=1}^m g_j(x)$
where each $g_j$ denotes a regression tree.  Thus, we approximate
(\ref{basemodel}) by a sum-of-trees model
\begin{equation}
Y = h(x) + \epsilon, \qquad  \epsilon
\sim N(0,\sigma^2). \label{sstmodel1}
\end{equation}

A sum-of-trees model is fundamentally an additive model with multivariate components.
Compared to generalized additive models based on sums of low dimensional smoothers,
these multivariate components can more naturally incorporate interaction effects.
And compared to a single tree model, the sum-of-trees can more easily
incorporate additive effects.

Various methods which combine a set of tree models, so called
ensemble methods, have attracted much attention. These include
boosting [\citeasnoun{FreSch1997}, \citeasnoun{Fri2001}], bagging [\citet{Br96a}] and random
forests [\citet{Bre2001}], each of which use
different techniques to fit a linear combination of trees. Boosting
fits a sequence of single trees, using each tree to fit data
variation not explained by earlier trees in the sequence. Bagging
and random forests use randomization to
create a large number of independent trees, and then reduce
prediction variance by averaging predictions across the trees. Yet
another approach that results in a linear combination of trees is
Bayesian model averaging applied to the posterior arising from a
Bayesian single-tree model as in \citeasnoun{ChipGeorMcCu1998a}
(hereafter CGM98), \citeasnoun{DeniMallSmit1998}, \citeasnoun{Blan2004} and
\citeasnoun{WuTjeWes2007}.  Such model averaging uses posterior
probabilities as weights for averaging the predictions from
individual trees.

In this paper we propose a Bayesian approach called BART (Bayesian
Additive Regression Trees) which uses a sum of trees to model or
approximate  $f(x) = E(Y | x)$. The
essential idea is to elaborate the sum-of-trees model
(\ref{sstmodel1}) by imposing a prior that regularizes the fit
by keeping the individual tree effects small.  In effect, the $g_j$'s become
a dimensionally adaptive random basis of ``weak learners,'' to
borrow a phrase from the boosting literature.  By weakening the $g_j$ effects,
BART ends up with a sum of trees, each of which explains a small and
different portion of $f$.  Note that BART is not equivalent to posterior averaging
of single tree fits of the entire function $f$.

To fit the sum-of-trees model, BART uses a tailored version of
Bayesian backfitting MCMC [\citet{HastTibs2000}] that
iteratively constructs and fits successive residuals. Although
similar in spirit to the gradient boosting approach of
\citeasnoun{Fri2001}, BART differs in both how it weakens the
individual trees by instead using a prior, and how it performs the
iterative fitting by instead using Bayesian backfitting on a fixed
number of trees.  Conceptually, BART can be viewed as a Bayesian
nonparametric approach that fits a parameter rich model using
a strongly influential prior distribution.

Inferences obtained from BART are based on successive iterations of the backfitting algorithm
which are effectively an MCMC sample from the induced posterior over the sum-of-trees model space.
A single posterior mean estimate of $f(x) = E(Y | x)$ at
any input value $x$ is obtained by a simple average of these successive
sum-of-trees model draws evaluated at $x$.  Further, pointwise uncertainty intervals for $f(x)$ are
easily obtained from the corresponding quantiles of the sample of draws.
Point and interval estimates are similarly obtained for functionals of $f$, such as partial dependence functions which reveal the marginal effects
of the $x$ components.
Finally, by keeping track of the relative frequency with
which $x$ components appear in the sum-of-trees model iterations, BART can be used to identify
which components are more important for explaining
the variation of $Y$.  Such variable selection information is model-free in the sense that it is not based on the usual assumption of an encompassing parametric model.

To facilitate the use of the BART methods described in this paper,
we have provided open-source software implementing BART as a stand-alone package or with an
interface to R, along with full documentation and examples.  It is
available as the \texttt{BayesTree} library in R at \url{http://cran.r-project.org/}.

The remainder of the paper is organized as follows.  In Section
\ref{sec:model} the BART model is outlined.  This consists of the
sum-of-trees model combined with a regularization prior. In
Section \ref{sec:postcalc} a Bayesian backfitting MCMC algorithm
and methods for inference are described. In Section \ref{sec:classification}
we describe a probit extension of BART for classification of binary $Y$.
In Section \ref{sec:examples} examples, both simulated and real, are used to
demonstrate the potential of BART. Section \ref{sec:executiontime} provides studies of execution time.
Section \ref{sec:related} describes extensions and
a variety of recent developments and applications of BART based on an early version of
this paper.  Section~\ref{sec:disc} concludes with a discussion.

\section{The BART model}\label{sec:model}

As described in the \hyperref[sec:intro]{Introduction}, the BART model consists of two parts: a sum-of-trees model
and a regularization prior on the parameters of that model.  We describe each of these in detail in the following subsections.

\subsection{A sum-of-trees model}

To elaborate the form of the sum-of-trees mod\-el~(\ref{sstmodel1}), we begin by
establishing notation for a single tree model. Let $T$ denote a
binary tree consisting of a set of interior node decision rules
and a set of terminal nodes, and let $M =
\{\mu_1,\mu_2,\ldots,\mu_b\}$ denote a set of parameter values
associated with each of the $b$ terminal nodes of $T$. The
decision rules are binary splits of the predictor space of the
form $\{x \in A\}$ vs $\{x \notin A\}$ where $A$ is a subset of
the range of~$x$.  These are typically based on the single components of
$x = (x_1,\ldots,x_p)$ and are of the
form $\{x_i \le c\}$ vs $\{x_i > c\}$ for continuous $x_i$.
Each $x$ value is associated with a single
terminal node of $T$ by the sequence of decision rules from top to
bottom, and is then assigned the $\mu_i$ value associated with
this terminal node. For a given $T$ and $M$, we use $g(x; T,M)$ to
denote the function which assigns a $\mu_i \in M$ to $x$.
Thus,
\begin{equation}\label{stmodel}
Y = g(x; T,M) + \epsilon,\qquad
   \epsilon \sim N(0,\sigma^2)
\end{equation}
is a single tree model of the form considered by CGM98.  Under (\ref{stmodel}),
the conditional mean of $Y$ given $x$, $E(Y | x)$ equals the
terminal node parameter $\mu_i$ assigned by $g(x;T,M)$.

With this notation, the sum-of-trees model (\ref{sstmodel1}) can
be more explicitly expressed as
\begin{equation}\label{sstmodel2}
Y = \sum_{j=1}^m g(x; T_j,M_j)   +
\epsilon,\qquad
   \epsilon \sim N(0,\sigma^2),
\end{equation}
where for each binary regression tree $T_j$ and its associated terminal node parameters $M_j$, $g(x; T_j,M_j)$ is the function which assigns $\mu_{ij} \in M_j$ to $x$.
Under (\ref{sstmodel2}), $E(Y | x)$  equals the sum of all the terminal node
$\mu_{ij}$'s assigned to $x$ by the $g(x; T_j,M_j)$'s.
When the number of trees $m > 1$,  each $\mu_{ij}$ here is merely a part of $E(Y | x)$, unlike the single tree model (\ref{stmodel}).  Furthermore, each such $\mu_{ij}$ will represent a main effect when $g(x;
T_j,M_j)$ depends on only one component of $x$ (i.e., a single
variable), and will represent an interaction effect when $g(x;
T_j,M_j)$ depends on more than one component of $x$ (i.e., more than
one variable).  Thus, the sum-of-trees model can incorporate both
main effects and interaction effects.  And because (\ref{sstmodel2})
may be based on trees of varying sizes, the interaction effects may
be of varying orders. In the special case where every terminal node
assignment depends on just a single component of $x$, the
sum-of-trees model reduces to a simple additive function, a sum of step functions of the
individual components of $x$.

With a large number of trees, a sum-of-trees model gains increased
representation flexibility which, as we'll see, endows BART with
excellent predictive capabilities.    This
representational flexibility is obtained by rapidly increasing the
number of parameters. Indeed, for fixed $m$, each sum-of-trees model
(\ref{sstmodel2}) is determined by $(T_1,M_1),\ldots,(T_m,M_m)$ and
$\sigma$, which includes all the bottom node parameters as well as
the tree structures and decision rules.  Further, the
representational flexibility of each individual tree leads to
substantial redundancy across the tree components. Indeed, one can
regard $\{g(x; T_1,M_1),
 \ldots , g(x; T_m,M_m)\}$ as an ``overcomplete
basis'' in the sense that many different choices of $(T_1,M_1),\ldots,(T_m,M_m)$ can lead to an identical function $\sum_{j=1}^m g(x; T_j,M_j)$.

\subsection{A regularization prior} \label{sec:prior}

We complete the BART model specification by imposing a prior over all
the parameters of the sum-of-trees model, namely,
$(T_1,M_1),\ldots,(T_m,M_m)$ and $\sigma$.   As discussed below, we
advocate specifications of this prior that effectively regularize the
fit  by keeping the individual tree effects from being unduly
influential.   Without such a regularizing influence, large tree
components would overwhelm the rich structure of (\ref{sstmodel2}),
thereby limiting the advantages of the additive representation both in
terms of function approximation and computation.

To facilitate the easy implementation of BART in practice, we recommend
automatic default specifications below which appear to be remarkably
effective, as demonstrated in the many examples of Section
\ref{sec:examples}.  Basically we proceed by first reducing the prior
formulation problem to the specification of just a few interpretable
hyperparameters which govern priors on $T_j$, $M_j$ and $\sigma$.  Our
recommended defaults are then obtained by using the observed variation
in $y$ to gauge reasonable hyperparameter values when external
subjective information is unavailable.  Alternatively, one can use the
considerations below to specify a range of plausible hyperparameter
values and then use cross-validation to select from these values.  This
will of course be computationally more demanding.  We should also
mention that although we sacrifice Bayesian coherence by using the data
to calibrate our priors, our overriding concern is to make sure that
our priors are not in severe conflict with the data.

\subsubsection{Prior independence and symmetry}\label{sec:simplification}

Specification of our regularization prior is vastly simplified by restricting attention to
priors for which
\begin{eqnarray}\label{indep1}
p((T_1,M_1),\ldots,(T_m,M_m),\sigma)
 & = &  \biggl[\prod_j p(T_j,M_j) \biggr]   p(\sigma)
 \nonumber\\[-8pt]\\[-8pt]
 & = &  \biggl[\prod_j p(M_j | T_j) p(T_j) \biggr]\nonumber
 p(\sigma)
\end{eqnarray}
and
\begin{equation}\label{indep2}
p(M_j   |   T_j) = \prod_i{p(\mu_{ij} | T_j)},
\end{equation}
where $\mu_{ij} \in M_j$.  Under such
priors, the tree components $(T_j,M_j)$ are independent of each other and of
 $\sigma$, and the terminal node parameters of every tree are
independent.

The independence restrictions above simplify the prior specification
problem to the specification of forms for just $p(T_j),
p(\mu_{ij} | T_j)$ and $p(\sigma)$, a specification which we further simplify by
using identical forms for all $p(T_j)$ and for all
$p(\mu_{ij} | T_j)$.  As described in the ensuing
subsections, for these we use the same prior forms proposed by CGM98 for
Bayesian CART.  In addition to their valuable computational
benefits, these forms are controlled by just a few interpretable
hyperparameters which can be calibrated using the data to yield
effective default specifications for regularization of the sum-of-trees model.
However, as will be seen, considerations for the choice of these hyperparameter
values for BART are markedly different than those for Bayesian CART.

\subsubsection{The $T_j$ prior}\label{sec:treeprior}

For $p(T_j)$, the form recommended by CGM98 is easy to specify and
dovetails nicely with calculations for the backfitting MCMC
algorithm described later in Section \ref{sec:mcmc}.  It is
specified by three aspects: (i) the probability that a node at
depth $d$ ($= 0, 1,2,\ldots$) is nonterminal, given by
\begin{equation}\label{treeprior}
\alpha (1+d)^{-\beta},\qquad
\alpha \in (0,1), \beta \in [0, \infty),
\end{equation}
(ii) the distribution on the splitting variable assignments at
each interior node, and (iii) the distribution on the splitting
rule assignment in each interior node, conditional on the splitting
variable. For (ii) and (iii) we use the simple defaults used by
CGM98, namely, the uniform prior on available variables for (ii)
and the uniform prior on the discrete set of available splitting
values for (iii).  Although not strictly coherent from the
Bayesian point of view, this last choice has the appeal of invariance
under monotone transformations of the splitting variables.

In a single tree model (i.e., $m =1$), a tree with many terminal
nodes may be needed to model a complicated structure.  However, for
a sum-of-trees model, especially with $m$ large, we want the
regularization prior to keep the individual tree components small.
In our examples in Section \ref{sec:examples}, we do so by using
$\alpha=0.95$ and $\beta=2$ in (\ref{treeprior}). With this choice,
trees with 1, 2, 3, 4 and $\geq $5 terminal nodes receive prior
probability of 0.05, 0.55, 0.28, 0.09 and 0.03, respectively.
Note that even with this prior, which puts most probability on
tree sizes of 2 or 3, trees with many terminal nodes can
be grown if the data demands it.  For example, in one of our
simulated examples with this prior, we observed considerable
posterior probability on trees of size 17 when we set $m=1$.

\subsubsection{The $\mu_{ij} | T_j$ prior}\label{sec:mprior}

For $p(\mu_{ij} | T_j)$, we use the conjugate normal distribution
$N(\mu_\mu, \sigma_\mu^2)$ which offers tremendous computational
benefits because $\mu_{ij}$ can be margined out.  To guide the
specification of the hyperparameters $\mu_\mu$ and $ \sigma_\mu$, note
that $E(Y | x)$ is the sum of $m$ $\mu_{ij}$'s under the sum-of-trees
model, and because the $\mu_{ij}$'s are apriori i.i.d., the induced
prior on $E(Y | x)$ is $N(m  \mu_\mu, m  \sigma_\mu^2)$. Note also that
it is highly probable that $E(Y | x)$ is between $y_{\min}$ and
$y_{\max}$, the observed minimum and maximum of $Y$ in the data. The
essence of our strategy is then to choose $\mu_\mu$ and $\sigma_\mu$ so
that $N(m   \mu_\mu, m  \sigma_\mu^2)$ assigns substantial probability
to the interval $(y_{\min}, y_{\max})$.  This can be conveniently done
by choosing $\mu_\mu$ and $ \sigma_\mu$ so that $m  \mu_\mu -  k
\sqrt{m}  \sigma_\mu = y_{\min}$ and $m   \mu_\mu +  k  \sqrt{m}
\sigma_\mu = y_{\max}$ for some preselected value of~$k$. For example,
$k=2$ would yield a 95\% prior probability that $E(Y | x)$ is in the
interval $(y_{\min}, y_{\max})$.

The strategy above uses an aspect of the observed data, namely, $y_{\min}$ and $y_{\max}$, to
try to ensure that the implicit prior for $E(Y | x)$ is in the right ``ballpark.''
That is to say, we want it to assign substantial probability to the entire region of
plausible values of $E(Y | x)$ while avoiding overconcentration and overdispersion.
We have found that, as long as this goal is met, BART is very robust to changes in the exact specification.
Such a data-informed prior approach is especially useful in our problem,  where reliable subjective information about $E(Y | x)$ is likely to be unavailable.

For convenience, we implement our specification strategy by first shifting and rescaling
$Y$ so that the observed transformed $y$ values range from  $y_{\min}= -0.5$ to $y_{\max}= 0.5$,
and then treating this transformed $Y$ as our dependent variable.
We then simply center the prior for $\mu_{ij}$ at zero $\mu_\mu
= 0$ and choose $\sigma_\mu$ so that
$ k \sqrt{m} \sigma_{\mu} = 0.5$ for
a suitable value of $k$,  yielding
\begin{equation}\label{eq:muprior}
\mu_{ij} \sim N(0,\sigma_{\mu}^2)\qquad \mbox{where     }
\sigma_{\mu} = 0.5/k \sqrt{m}.
\end{equation}

This prior has the effect of shrinking the tree parameters
$\mu_{ij}$ toward zero, limiting the effect of the individual
tree components of (\ref{sstmodel2}) by keeping them small.  Note
that as $k$ and/or the number of trees $m$ is increased, this
prior will become tighter and apply greater shrinkage to the
$\mu_{ij}$'s.  Prior shrinkage on the $\mu_{ij}$'s is the
counterpart of the shrinkage parameter in
Friedman's (\citeyear{Fri2001}) gradient boosting algorithm.  The prior
standard deviation $\sigma_{\mu}$ of $\mu_{ij}$  here and the gradient
boosting shrinkage parameter there both serve to ``weaken'' the
individual trees so that each is constrained to play a smaller
role in the overall fit.  For the choice of $k$, we have found
that values of $k$ between 1 and 3 yield good results, and we
recommend $k = 2$ as an automatic default choice. Alternatively,
the value of $k$ may be chosen by cross-validation from a range of
reasonable choices.

Although the calibration of this prior is based on a simple linear
transformation of $Y$, it should be noted that there is no need to
transform the predictor variables.  This is a consequence of the
fact that the tree splitting rules are invariant to monotone
transformations of the $x$ components.   The simplicity of
our prior for $\mu_{ij}$ is an appealing feature of BART.
In contrast, methods like neural nets that use linear combinations
of predictors require standardization choices for each predictor.

\subsubsection{The $\sigma$ prior}\label{sec:sigmaprior}

For $p(\sigma)$, we also use a conjugate prior, here the inverse chi-square distribution
$\sigma^2 \sim \nu   \lambda/\chi_{\nu}^2.$  To guide the specification
of the hyperparameters $\nu$ and $\lambda$, we again use a data-informed prior approach,
in this case to assign substantial probability to the entire region of
plausible values of $\sigma$ while avoiding overconcentration and overdispersion.
Essentially, we calibrate the prior df $\nu$ and scale $\lambda$ for this purpose
using a ``rough data-based overestimate''
$\hat{\sigma}$ of $\sigma$.  Two natural choices for $\hat{\sigma}$ are (1) the ``naive''
specification, in which we take $\hat{\sigma}$ to be the sample
standard deviation of $Y$, or (2) the ``linear model''
specification, in which we take $\hat{\sigma}$ as the residual
standard deviation from a least squares linear regression of $Y$
on the original $X$'s.  We then pick a value of
$\nu$ between 3 and 10 to get an appropriate shape, and a value of
$\lambda$ so that the $q$th quantile of the prior on $\sigma$ is located at
$\hat{\sigma}$, that is, $P(\sigma < \hat{\sigma}) = q.$  We consider values of $q$ such as
0.75, 0.90 or 0.99 to center the distribution below $\hat{\sigma}$.

\begin{figure}

\includegraphics{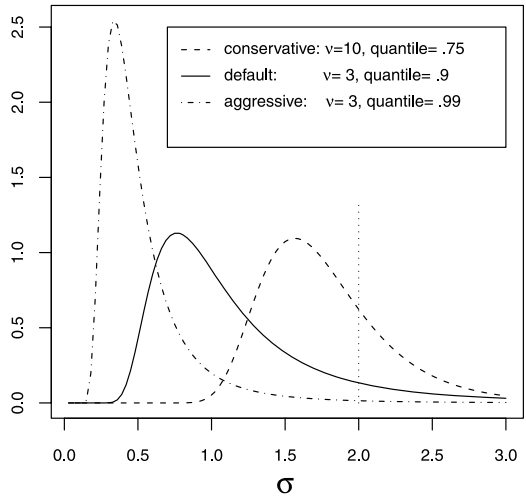}

\caption{Three priors on $\sigma$ based on $\mathrm{df} = \nu$ and $\mathrm{quantile} = q$
when $\hat \sigma = 2$.}\label{fig:sigmaprior}
\end{figure}

Figure \ref{fig:sigmaprior} illustrates priors corresponding to
three $(\nu, q)$ settings when the rough overestimate is
$\hat{\sigma}=2$.  We refer to these three settings, $(\nu, q) =
(10, 0.75)$, $(3, 0.90)$, $(3, 0.99)$, as conservative, default
and aggressive, respectively.  The prior mode moves toward
smaller $\sigma$ values as $q$ is increased.  We recommend against
choosing $\nu < 3$ because it seems to concentrate too much mass
on very small $\sigma$ values, which leads to overfitting. In our
examples, we have found these three settings to work very well and
yield similar results.  For automatic use, we recommend the
default setting $(\nu, q) = (3, 0.90)$ which tends to avoid
extremes.  Alternatively, the values of $(\nu, q)$ may be chosen by
cross-validation from a range of
reasonable choices.

\subsubsection{The choice of $m$}\label{sec:numtrees}

A major difference between BART and  boosting methods is that for a
fixed number of trees $m$,  BART uses an iterative backfitting
algorithm (described in Section \ref{sec:mcmc}) to cycle over and over
through the $m$ trees.  If BART is to be used for estimating $f(x)$ or
predicting $Y$, it might be reasonable to treat  $m$ as an unknown
parameter, putting a prior on $m$ and proceeding with a fully Bayes
implementation of BART.  Another reasonable strategy might be to select
a ``best'' value for $m$ by cross-validation from a range of reasonable
choices.  However, both of these strategies substantially increase
computational requirements.

To avoid the computational costs of these strategies, we have found it
fast and expedient for estimation and prediction to begin with a
default of $m = 200$, and then perhaps to check if one or two other
choices makes any difference.  Our experience has been that as $m$ is
increased, starting with $m = 1$, the predictive performance of BART
improves dramatically until at some point it levels off  and then
begins to very slowly degrade for large values of $m$.   Thus, for
prediction, it seems only important to avoid choosing $m$ too small.
As will be seen in Section \ref{sec:examples},  BART yielded excellent
predictive performance on a wide variety of examples with the simple
default $m = 200$.  Finally, as we shall see later in Sections
\ref{sec:estpd} and \ref{sec:examples}, other considerations for
choosing $m$ come into play when BART is used for variable selection.

\section{Extracting information from the posterior}\label{sec:postcalc}

\subsection{A Bayesian backfitting MCMC algorithm} \label{sec:mcmc}

Given the observed data $y$, our Bayesian setup induces a
posterior distribution
\begin{equation} \label{posterior}
p((T_1,M_1), \ldots,(T_m,M_m),\sigma |  y)
\end{equation}
on all the unknowns that determine a sum-of-trees model
(\ref{sstmodel2}).  Although the sheer size of the parameter
space precludes exhaustive calculation,  the following backfitting
MCMC algorithm can be used to sample from this posterior.

At a general level, our algorithm is a Gibbs sampler.  For
notational convenience, let $T_{(j)}$ be the set of all trees in
the sum \textit{except} $T_j$, and similarly define $M_{(j)}$. Thus,
$T_{(j)}$ will be a set of $m-1$ trees, and $M_{(j)}$ the
associated terminal node parameters.  The Gibbs sampler here
entails $m$ successive draws of $(T_j,M_j)$ conditionally on
$(T_{(j)}, M_{(j)}, \sigma)$:
\begin{equation}\label{draw1}
(T_j,M_j) | T_{(j)}, M_{(j)}, \sigma, y,
\end{equation}
$j = 1,\ldots,m$, followed by a draw of $\sigma$ from the full conditional:
\begin{equation}\label{draw2}
\sigma | T_1,  \ldots, T_m, M_1, \ldots, M_m, y  .
\end{equation}
\citeasnoun{HastTibs2000} considered a similar application of
the Gibbs sampler for posterior sampling for additive and
generalized additive models with $\sigma$ fixed, and showed how it
was a stochastic generalization of the backfitting algorithm for
such models.  For this reason, we refer to our algorithm as
backfitting MCMC.

The draw of $\sigma$ in (\ref{draw2}) is simply a draw from an
inverse gamma distribution and so can be easily obtained by
routine methods.  More challenging is how to implement the
$m$ draws of $(T_j,M_j)$ in (\ref{draw1}). This can be done by
taking advantage of the following reductions. First, observe that
the conditional distribution $p(T_j,M_j |   T_{(j)}, M_{(j)},
\sigma , y)$ depends on $(T_{(j)}, M_{(j)}, y)$ only through
\begin{equation}
R_j \equiv y - \sum_{k \neq j} g(x;T_k,M_k),
\end{equation}
the $n$-vector of partial residuals based on a fit that excludes
the $j$th tree. Thus, the $m$ draws of $(T_j,M_j)$ given $(T_{(j)},
M_{(j)}, \sigma , y)$ in (\ref{draw1}) are equivalent to $m$ draws
from
\begin{equation} \label{newdraw}
(T_j,M_j) | R_j, \sigma,
\end{equation}
$j = 1,\ldots,m$.

Now (\ref{newdraw}) is formally equivalent to the posterior of the
single tree model  $R_j = g(x; T_j,M_j) + \epsilon$ where $R_j$
plays the role of the data $y$.  Because we have used a conjugate
prior for $M_j$,
\begin{equation}\label{Mmarg}
p(T_j | R_j,\sigma) \propto p(T_j) \int p(R_j | M_j,T_j, \sigma)
p(M_j|T_j,\sigma) \,dM_j
\end{equation}
can be obtained in closed form up to a norming constant.  This
allows us to carry out each draw from (\ref{newdraw})
in two successive steps as
\begin{eqnarray} \label{tdraw}
&T_j | R_j,\sigma,&
\\ \label{mdraw}
&M_j| T_j, R_j, \sigma .&
\end{eqnarray}

The draw of $T_j$ in (\ref{tdraw}), although somewhat elaborate,
can be obtained using the Metropolis--Hastings (MH) algorithm of
CGM98. This algorithm proposes a new tree based on the current
tree using one of four moves.  The moves and their associated
proposal probabilities are as follows: growing a terminal node (0.25),
pruning a pair of terminal nodes (0.25), changing a nonterminal
rule (0.40), and swapping a rule between parent and child (0.10).
Although the grow and prune moves change the number of
terminal nodes,
by integrating out $M_j$ in (\ref{Mmarg}), we
avoid the complexities associated with reversible jumps between
continuous spaces of varying dimensions [\citet{Gre1995}].

Finally, the draw of $M_j$ in (\ref{mdraw}) is simply a set of
independent draws of the terminal node $\mu_{ij}$'s from a normal
distribution. The draw of $M_j$ enables the calculation of the
subsequent residual $R_{j+1}$ which is critical for the next draw
of $T_j$. Fortunately, there is again no need for a complex
reversible jump implementation.

We initialize the chain with $m$ simple single node trees, and then
iterations are repeated until satisfactory convergence is
obtained.  At each iteration, each tree may increase or decrease
the number of terminal nodes by one, or change one or two decision
rules.  Each $\mu$ will change (or cease to exist or be born), and
$\sigma$ will change. It is not uncommon for a tree to grow large
and then subsequently collapse back down to a single node as the
algorithm iterates. The sum-of-trees model, with its abundance of
unidentified parameters, allows for ``fit'' to be freely
reallocated from one tree to another. Because each move makes only
small incremental changes to the fit, we can imagine the algorithm
as analogous to sculpting a complex figure by adding and
subtracting small dabs of clay.

Compared to the single tree model MCMC approach of CGM98, our
backfitting MCMC algorithm mixes dramatically better.  When only
single tree models are considered, the MCMC algorithm tends to
quickly gravitate toward a single large tree and then gets stuck
in a local neighborhood of that tree. In sharp contrast, we have
found that restarts of the backfitting MCMC algorithm give
remarkably similar results even in difficult problems.
Consequently, we run one long chain with BART rather than multiple
starts.  Although mixing does not appear to be an issue, the
recently proposed modifications of \citeasnoun{Blan2004} and \citeasnoun{WuTjeWes2007}
might well provide additional benefits.

\subsection{Posterior inference statistics}\label{sec:estpd}

The backfitting algorithm described in the previous section is ergodic,
generating a sequence of draws of $(T_1,M_1),\break \ldots,(T_m, M_m),\sigma$
which is converging (in distribution) to the posterior\break $p((T_1,M_1),
\ldots,(T_m, M_m),\sigma |  y)$.  The induced sequence of sum-of-trees
functions
\begin{equation}\label{fstar}
f^*(\cdot) = \sum_{j=1}^m g(\cdot;T_j^*,M_j^*),
\end{equation}
for the sequence of draws $(T_1^*,M_1^*), \ldots,(T_m^*,M_m^*)$, is
thus converging to $p(f   |  y)$, the posterior distribution on the
``true'' $f(\cdot)$.  Thus, by running the algorithm long enough after
a suitable burn-in period, the sequence of $f^*$ draws, say,
$f^*_1,\dots,f^*_K$, may be regarded as an approximate, dependent
sample of size $K$ from $p(f   |  y)$.  Bayesian inferential quantities
of interest can then be approximated with this sample as indicated
below.  Although the number of iterations needed for reliable
inferences will of course depend on the particular application, our
experience with the examples in Section \ref{sec:examples} suggests
that the number of iterations required is relatively modest.

To estimate $f(x)$ or predict $Y$ at a particular $x$, in-sample or
out-of-sample, a~natural choice is the average of the after burn-in
sample  $f^*_1,\dots,f^*_K$,
\begin{equation}\label{eq:fhat}
\frac{1}{K} \sum_{k=1}^K f^*_k(x),
\end{equation}
which approximates the posterior mean $E( f(x)  |  y)$.    Another good
choice would be the median of $f^*_1(x),\dots,f^*_K(x)$ which
approximates the posterior median of $f(x)$.  Posterior uncertainty
about $f(x)$ may be gauged by the variation of
$f^*_1(x),\dots,f^*_K(x)$. For example, a natural and convenient
$(1-\alpha)\%$ posterior interval for $f(x)$ is obtained as the
interval between the upper and lower $\alpha/2$ quantiles of
$f^*_1(x),\dots,f^*_K(x)$. As will be seen, these uncertainty intervals
behave sensibly, for example, by widening at $x$ values far from the
data.

It is also straightforward to use $f^*_1(x),\dots,f^*_K(x)$ to  estimate
other functionals of $f$.  For example, a functional of particular interest is the
partial dependence function [\citet{Fri2001}], which summarizes the
marginal effect of one (or more) predictors on the response.  More precisely,
letting $f(x) = f(x_s,x_c)$ where $x$ has been partitioned into the predictors of interest, $x_s$
and the complement $x_c=x\setminus x_s$, the partial dependence function is defined as
\begin{equation}\label{eq:pdsum}
f(x_s) = \frac{1}{n}\sum_{i=1}^n f(x_s,x_{ic}),
\end{equation}
where $x_{ic}$ is the $i$th observation of $x_c$ in the data. Note
that $(x_s,x_{ic})$ will not generally be one of the observed data
points.  A draw from the induced BART posterior $p(f(x_s) |  y)$ at any value of $x_s$ is obtained by
simply computing $f^*_k(x_s)= \frac{1}{n}\sum_i f^*_k(x_s,x_{ic})$.  The average of $f^*_1(x_s),\dots,f^*_K(x_s)$ then yields an
estimate of $f(x_s)$, and the upper and lower $\alpha/2$ quantiles  provide endpoints
of $(1-\alpha)\%$ posterior intervals for $f(x_s)$.

Finally, as mentioned in Section \ref{sec:intro}, BART can also be used
for variable selection by selecting those variables that appear most
often in the fitted sum-of-trees models.  Interestingly, this strategy
is less effective when $m$ is large because the redundancy offered by
so many trees tends to mix many irrelevant predictors in with the
relevant ones.  However, as $m$ is decreased and that redundancy is
diminished, BART tends to heavily favor relevant predictors for its
fit.  In a sense, when $m$ is small the predictors compete with each
other to improve the fit.

This model-free approach to variable selection is accomplished by
observing what happens to the $x$ component usage frequencies in a
sequence of MCMC samples $f^*_1,\dots,f^*_K$ as the number of trees $m$
is set smaller and smaller.  More precisely, for each simulated
sum-of-trees model $f^*_k$, let $z_{ik}$ be the proportion of all
splitting rules that use the $i$th component of $x$.  Then
\begin{equation}\label{eq:xfreq}
v_i \equiv \frac {1}{K} \sum_{k=1}^K z_{ik}
\end{equation}
is the average use per splitting rule for the $i$th component of $x$.
As $m$ is set smaller and smaller, the sum-of-trees models tend to more
strongly favor inclusion of those $x$ components which improve
prediction of $y$ and exclusion of those $x$ components that are
unrelated to $y$.  In effect,  smaller $m$ seems to create a bottleneck
that forces the $x$ components to compete for entry into the
sum-of-trees model.  As we shall see illustrated in Section
\ref{sec:examples}, the $x$ components with the larger $v_i$'s will
then be those that provide the most information for predicting $y$.
Finally, it might be useful to consider alternative ways of measuring
component usage in (\ref{eq:xfreq}) such as weighting variables by the
number of data points present in the node, thereby giving more weight
to the importance of initial node splits.

\section{BART probit for classification}\label{sec:classification}

Our development of BART up to this point has pertained to setups where
the output of interest $Y$ is a continuous variable.  However, for
binary $Y$ ($= 0$ or 1), it is straightforward to extend BART to the
probit model setup for classification
\begin{equation}\label{eq:cmodel}
p(x) \equiv P[Y = 1  | x] = \Phi[G(x)],
\end{equation}
where
\begin{equation}
G(x) \equiv \sum_{j=1}^m g(x; T_j,M_j)
\end{equation}
and $\Phi[\cdot]$ is the standard normal c.d.f.  Note that each
classification probability $p(x)$ here is obtained as a function of
$G(x)$, our sum of regression trees.  This contrasts with the often
used aggregate classifier approaches which use a majority or an average
vote based on an ensemble of classification trees, for example, see
\citeasnoun{AmitGema1997} and \citeasnoun{Bre2001}.

For the BART extension to (\ref{eq:cmodel}), we need to impose a
regularization prior on $G(x)$ and to implement a Bayesian backfitting
algorithm for posterior computation.  Fortunately, these are obtained
with only minor modifications of the methods in Sections
\ref{sec:model} and \ref{sec:postcalc}. As opposed to
(\ref{sstmodel2}), the model (\ref{eq:cmodel}) implicitly assumes
$\sigma = 1$ and so only a prior on $(T_1,M_1),\ldots,(T_m,M_m)$ is
needed.  Proceeding exactly as in Section \ref{sec:simplification}, we
consider a prior of the form
\begin{equation}
p((T_1,M_1),\ldots,(T_m,M_m))= \prod_j  \biggl[p(T_j) \prod_i p(\mu_{ij} | T_j)  \biggr],
\end{equation}
where each tree prior $p(T_j)$ is the choice recommended in  Section
\ref{sec:treeprior}.   For the choice of $p(\mu_{ij} | T_j)$ here, we
consider the case where the interval $(\Phi[-3.0], \Phi[3.0])$ contains
most of the $p(x)$ values of interest, a case which will often be of
practical relevance.  Proceeding similarly to the motivation of
(\ref{eq:muprior}) in Section \ref{sec:mprior}, we would then recommend
the choice
\begin{equation}\label{eq:muprior2}
\mu_{ij} \sim N(0,\sigma_{\mu}^2)\qquad \mbox{where     }
\sigma_{\mu} = 3.0/k \sqrt{m} ,
\end{equation}
where $k$ is such that $G(x)$ will with high probability be in the interval $(-3.0,3.0)$.
Just as for (\ref{eq:muprior}), this prior has the effect of shrinking the tree parameters
$\mu_{ij}$ toward zero, limiting the effect of the individual
tree components of $G(x)$.  As $k$ and/or the number of trees $m$ is increased, this
prior will become tighter and apply greater shrinkage to the
$\mu_{ij}$'s.  For the choice of $k$, we have found
that values of $k$ between 1 and 3 yield good results, and we
recommend $k = 2$ as an automatic default choice. Alternatively,
the value of $k$ may be chosen by cross-validation.

By shrinking $G(x)$ toward 0, the prior (\ref{eq:muprior2}) has the effect of shrinking
$p(x) = \Phi[G(x)]$ toward $0.5$.  If it is of interest to shrink toward
a value $p_0$ other than $0.5$, one can simply replace $G(x)$ by $G_c = G(x)+ c$ in
(\ref{eq:cmodel}) with the offset $c = \Phi^{-1}[p_0]$.  Note also that if an interval other than $(\Phi[-3.0], \Phi[3.0])$
is of interest for $p(x)$, suitable modification of (\ref{eq:muprior2}) is straightforward.

Turning to posterior calculation, the essential features of the
backfitting algorithm in Section \ref{sec:mcmc} can be implemented by
using the augmentation idea of \citeasnoun{AlbeChib1993}. The key idea
is to recast the model (\ref{eq:cmodel}) by introducing latent
variables $Z_1,\ldots,Z_n$ i.i.d. $\sim N(G(x), 1)$ such that $Y_i = 1$
if $Z_i > 0$ and $Y_i = 0$ if $Z_i \le 0$.  Note that under this
formulation, $Z_i  |  [y_i = 1] \sim \max\{N(g(x), 1), 0\}$ and $Z_i
|  [y_i = 0] \sim \min\{N(g(x), 1), 0\}$.  Incorporating simulation of
the latent $Z_i$ values into the backfitting algorithm, the Gibbs
sampler iterations here entail $n$ successive draws of $Z_i  |  y_i$,
$i = 1,\dots, n$,  followed by $m$ successive draws of $(T_j,M_j) |T_{(j)}, M_{(j)}, z_1,\ldots,z_n$, $j = 1,\ldots,m$, as spelled out in
Section \ref{sec:mcmc}.  The induced sequence of sum-of-trees functions
\begin{equation}\label{pstar}
p^*(\cdot) = \Phi \Biggl[\sum_{j=1}^m g(\cdot;T_j^*,M_j^*) \Biggr],
\end{equation}
for the sequence of draws $(T_1^*,M_1^*), \ldots,(T_m^*,M_m^*)$, is
thus converging to the posterior distribution on the ``true''
$p(\cdot)$.  After a suitable burn-in period, the sequence of $g^*$
draws, say, $g^*_1,\dots,g^*_K$, may be regarded as an approximate,
dependent sample from this posterior which can be used to draw
inference about $p(\cdot)$ in the same way that $f^*_1,\dots,f^*_K$ was
used in Section \ref{sec:estpd} to draw inference about $f(\cdot)$.

\section{Applications}\label{sec:examples}

In this section we demonstrate the application of BART on several
examples. We begin in Section \ref{sec:bakeoff} with a predictive
cross-validation performance comparison of BART with competing methods
on 42 different real data sets.  We next, in Section
\ref{sec:simex:friedman}, evaluate and illustrate BART's capabilities
on simulated data used by \citeasnoun{Frie1991}.  Finally, in Section
\ref{sec:drugdisc} we apply the BART probit model to a drug discovery
classification problem.  All of the BART calculations throughout this
section can be reproduced with the \texttt{BayesTree} library at
\url{http://cran.r-project.org/}.

\subsection{Predictive comparisons on 42 data sets}\label{sec:bakeoff}

Our first illustration is a ``bake-off,'' a predictive
performance comparison of BART with competing methods on 42 different real data sets.
These data sets (see Table \ref{tab:datasets}) are a subset of 52 sets considered by \citeasnoun{KimLohShiCha2007}.
Ten data sets were excluded either because Random Forests was unable to use over 32
categorical predictors, or because a single train/test split was used
in the original paper.  All data sets correspond to regression setups with
between 3 and 28 numeric predictors and 0 to 6 categorical predictors.
Categorical predictors were converted into 0/1 indicator variables
corresponding to each level.  Sample sizes vary from 96 to 6806 observations.
In each of the 42 data sets, the response was minimally preprocessed,
applying a log or square root transformation if this made the histogram
of observed responses more bell-shaped.  In about half the cases, a log
transform was used to reduce a right tail.  In one case (Fishery) a square
root transform was most appropriate.

\begin{table}
\tabcolsep=0pt
\caption{The 42 data sets used in the bake-off}\label{tab:datasets}
\begin{tabular*}{\textwidth}{@{\extracolsep{\fill}}lc@{\qquad}lc@{\qquad}lc@{\qquad}lc@{\qquad}lc@{}}
\hline
\textbf{Name} & $\bolds{n}$ &
\textbf{Name} & $\bolds{n}$ &
\textbf{Name} & $\bolds{n}$ &
\textbf{Name} & $\bolds{n}$ &
\textbf{Name} & $\bolds{n}$ \\
\hline
Abalone  &  4177  &  Budget  &  1729  &  Diamond  &  \hphantom{0}308  &  Labor     &  2953  &  Rate     &  144 \\
Ais      &  \hphantom{0}202   &  Cane    &  3775  &  Edu      &  1400 &  Laheart   &  \hphantom{0}200   &  Rice     &  171 \\
Alcohol  &  2462  &  Cardio  &  \hphantom{0}375   &  Enroll   &  \hphantom{0}258  &  Medicare  &  4406  &  Scenic   &  113 \\
Amenity  &  3044  &  College &  \hphantom{0}694   &  Fame     &  1318 &  Mpg       &  \hphantom{0}392   &  Servo    &  167 \\
Attend   &  \hphantom{0}838   &  Cps     &  \hphantom{0}534   &  Fat      &  \hphantom{0}252  &  Mumps     &  1523  &  Smsa     &  141 \\
Baseball &  \hphantom{0}263   &  Cpu     &  \hphantom{0}209   &  Fishery  &  6806 &  Mussels   &  \hphantom{0}201   &  Strike   &  625 \\
Baskball &  \hphantom{00}96    &  Deer    &  \hphantom{0}654   &  Hatco    &  \hphantom{0}100  &  Ozone     &  \hphantom{0}330   &  Tecator  &  215 \\
Boston   &  \hphantom{0}506   &  Diabetes&  \hphantom{0}375   &  Insur    &  2182 &  Price     &  \hphantom{0}159   &  Tree     &  100 \\
Edu      &  1400  &  Fame    &  1318  & \\
\hline
\end{tabular*}
\end{table}

For each of the 42 data sets, we created 20 independent train/test splits by
randomly selecting $5/6$ of the data as a training set and
the remaining $1/6$ as a test set.  Thus, $42 \times 20 = 840$
test/train splits were created.  Based on each training set,
each method was then used to predict the corresponding test set
and evaluated on the basis of its predictive RMSE.

We considered two versions of BART: BART-cv where the prior hyperparameters
$(\nu, q, k, m)$ were treated as operational parameters to
be tuned via cross-validation, and BART-default where
we set $(\nu, q, k, m)$ to the defaults $(3, 0.90, 2, 200)$.
For both BART-cv and BART-default, all specifications of the quantile
$q$ were made relative to the least squares linear regression estimate
$\hat{\sigma}$,  and the number of burn-in steps and MCMC iterations used
were determined by inspection of a single long run.  Typically, 200 burn-in
steps and 1000 iterations were used.  For BART prediction at each $x$, we used the
posterior mean estimates given by (\ref{eq:fhat}).

As competitors, we considered linear regression with L1 regularization
(the Lasso) [\citet{EfrHasJohTib2004}] and three black-box models:
gradient boosting [\citet{Fri2001}, implemented as \texttt{gbm} in
R by \citeasnoun{Rid2004}], random forests [\citet{Bre2001},
implemented as \texttt{randomforest} in R] and neural networks with one
layer of hidden units [implemented as \texttt{nnet} in R by
\citeasnoun{VenaRipl2002}]. These competitors were chosen because, like
BART, they are black box predictors. Trees, Bayesian CART (CGM98) and
Bayesian treed regression [\citet{ChipGeorMcCu2002a}] models were not
considered, since they tend to sacrifice predictive performance for
interpretability.

With the exception of BART-default (which requires no tuning),
the operational parameters
of every method were chosen via 5-fold cross-validation
within each training set.  The parameters considered and potential
levels are given in Table~\ref{tab:parameters}.   In particular, for BART-cv, we
considered the following:
\begin{itemize}
\item three settings $(3,0.90)$ (default), $(3,0.99)$ (aggressive) and
$(10,0.75)$ (conservative) as shown in Figure \ref{fig:sigmaprior}
for the $\sigma$ prior hyperparameters  $(\nu, q)$,
\item four values $k= 1,2,3,5$ reflecting moderate to heavy shrinkage for the $\mu$ prior hyperparameter, and
\item two values $m = 50, 200$ for the number of trees,
\end{itemize}
a total of $3*4*2 = 24$ potential choices for $(\nu, q, k, m)$.

\begin{table}[b]
\caption{Operational parameters for the various competing models}\label{tab:parameters}
\begin{tabular*}{\textwidth}{@{\extracolsep{\fill}}llc@{}}
\hline
 \textbf{Method} & \textbf{Parameter} & \textbf{Values considered} \\
 \hline
 BART-cv & Sigma prior: $(\nu, q)$
combinations & (3, 0.90), (3, 0.99), (10, 0.75)\\
& \# trees $m$ & 50, 200 \\
     & $\mu$ prior:  $k$ value for $\sigma_\mu$
          &  1, 2, 3, 5\\[3pt]
Lasso & Shrinkage (in range 0--1) & $0.1, 0.2, \ldots, 1.0$ \\[3pt]
Gradient boosting  & \# of trees & 50, 100, 200 \\
 & Shrinkage (multiplier of each tree added)& 0.01, 0.05, 0.10, 0.25 \\
         & Max depth permitted for each tree & 1, 2, 3, 4 \\[3pt]
Neural nets & \# hidden units &  see text \\
   & Weight decay & 0.0001, 0.001, 0.01, 0.1, 1, 2, 3\\[3pt] 
Random  forests    & \# of trees &   500 \\
    & \% variables sampled to grow each node &  10, 25, 50, 100\\
\hline
\end{tabular*}
\end{table}

All the levels in Table~\ref{tab:parameters} were chosen with a
sufficiently wide range so that the selected value was not at an extreme of
the candidate values in most problems.  Neural networks are the only model
whose operational parameters need additional explanation.  In that case,
the number of hidden units was chosen in terms of the implied number
of weights, rather than the number of units.  This design choice was
made because of the widely varying number of predictors across problems,
which directly impacts the number of weights.  A number of hidden units
were chosen so that there was a total of roughly $u$ weights, with $u=
50, 100, 200, 500$ or $800$.
In all cases, the number of hidden units was further constrained to fall
between 3 and~30.
For example, with 20 predictors we used 3, 8 and 21 as candidate
values for the number of hidden units.

\begin{figure}[b]

\includegraphics{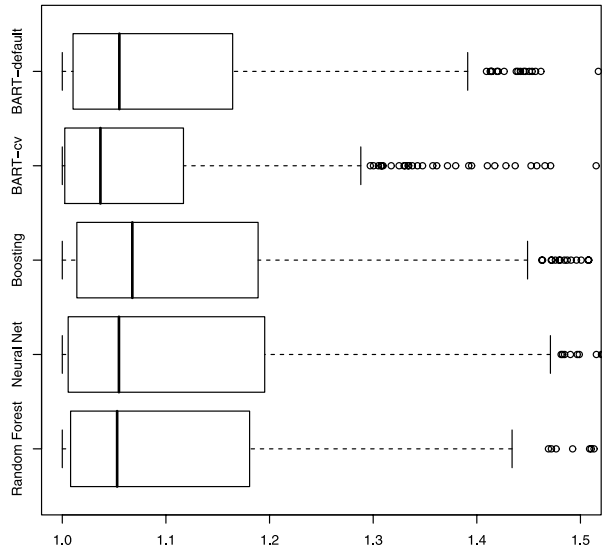}

\caption{Boxplots of the RRMSE values for each method across the 840 test/train splits.
Percentage RRMSE  values larger than 1.5 for each method (and not plotted)
were the following: random forests 16.2\%, neural net 9.0\%, boosting 13.6\%, BART-cv
9.0\% and BART-default 11.8\%.  The Lasso (not plotted because of too many
large RRMSE values) had 29.5\% greater than 1.5.}\label{fig:boxplot}
\end{figure}

To facilitate performance comparisons across data sets, we considered
relative RMSE (RRMSE), which we defined as the RMSE divided by the
minimum RMSE obtained by any method for each test/train split.  Thus, a
method obtained an RRMSE of 1.0 when that method had the minimum RMSE
on that split.  As opposed to the RMSE, the RRMSE provides meaningful
comparisons across data sets because of its invariance to location and
scale transformations of the response variables.  Boxplots of the 840
test/train split RRMSE values for each method are shown in
Figure~\ref{fig:boxplot}, and the (50\%, 75\%) RRMSE quantiles (the
center and rightmost edge of each box in Figure~\ref{fig:boxplot}) are
given in Table~\ref{tab:perf}.  (The Lasso was left off the boxplots
because its many large RRMSE values visually overwhelmed the other
comparisons.)

\begin{table}
\tablewidth=170pt
\caption{(50\%, 75\%) quantiles of relative RMSE values for each method across the 840 test/train splits}\label{tab:perf}
\begin{tabular*}{170pt}{@{\extracolsep{\fill}}lc@{}}
\hline
\textbf{Method}         &   \textbf{(50\%, 75\%)}     \\
\hline
Lasso         & (1.196, 1.762)\\
Boosting      & (1.068, 1.189)\\
Neural net    & (1.055, 1.195)\\
Random forest & (1.053, 1.181)\\
BART-default  & (1.055, 1.164)\\
BART-cv       & (1.037, 1.117)\\
\hline
\end{tabular*}
\end{table}

Although relative performance in Figure~\ref{fig:boxplot} varies widely across the different problems, it is clear from the distribution of RRMSE values  that BART-cv tended to more often obtain smaller RMSE than any of its competitors.   Also notable is the overall performance of BART-default which was arguably second best.  This is especially impressive since neural nets, random forests and gradient boosting all relied here on cross-validation for control parameter tuning.   By avoiding the need for hyperparameter specification, BART-default is vastly easier and faster to use. For example, a single implementation of BART-cv here requires
selection among the 24 possible hyperparameter values with 5 fold cv,
followed by fitting the best model, for a total of $24*5 + 1 = 121$
applications of BART.  For those who want a computationally inexpensive method ready for
easy ``off the shelf'' use, BART-default is the winner in this experiment.

\subsection{Friedman's five dimensional test function}\label{sec:simex:friedman}

We next proceed to illustrate various features of BART on simulated data where
we can gauge its performance against the true underlying signal. For this purpose,
we  constructed  data by
simulating values of $x = (x_1,x_2,\ldots,x_p)$ where
\begin{equation}\label{eq:fri-xs}
x_1,x_2,\ldots,x_p \mbox{  i.i.d. } \sim \operatorname{Uniform}(0,1),
\end{equation}
and $y$ given $x$ where
\begin{equation}\label{eq:fri-ys}
y = f(x) + \epsilon = 10 \sin(\pi x_1 x_2) + 20 (x_3-0.5)^2 + 10
x_4 + 5 x_5 + \epsilon,
\end{equation}
where $\epsilon \sim N(0,1)$.  Because $y$ only depends on
$x_1,\ldots,x_5$, the predictors $x_6,\ldots, x_p$ are irrelevant.
These added variables together with the interactions and
nonlinearities make it more challenging to find $f(x)$ by
standard parametric methods.  \citeasnoun{Frie1991} used this setup
with $p = 10$ to illustrate the potential of
multivariate adaptive regression
splines (MARS).

In Section \ref{sec:friedman-simple} we illustrate various basic
features of BART.  We illustrate point and interval estimation of
$f(x)$,  model-free variable selection and estimation of partial
dependence functions.  We see that the BART MCMC burns-in quickly and
mixes well. We illustrate BART's robust performance with respect to
various hyperparameter settings.  In Section \ref{sec:friedman-finding}
we increase the number of irrelevant predictors in the data to show
BART's effectiveness at detecting a low dimensional structure in a high
dimensional setup. In Section \ref{sec:friedman-train-test} we compare
BART's out-of-sample performance with the same set of competitors used
in Section \ref{sec:bakeoff} with $p$ equal to 10, 100 and 1000. We
find that BART dramatically outperforms the other methods.

\subsubsection{A simple application of BART}\label{sec:friedman-simple}

We begin by illustrating the basic features of BART on a single simulated data set
of the Friedman function (\ref{eq:fri-xs}) and (\ref{eq:fri-ys})
with $p= 10$ $x's$ and $n = 100$ observations. For simplicity, we
applied BART with the default setting $(\nu, q, k,m) = (3,0.90,2,200)$
described in Section \ref{sec:prior}.
Using the backfitting MCMC algorithm, we generated 5000 MCMC draws
of $f^*$ as in (\ref{fstar}) from the posterior  after skipping
1000 burn-in iterations.

\begin{figure}[b]

\includegraphics{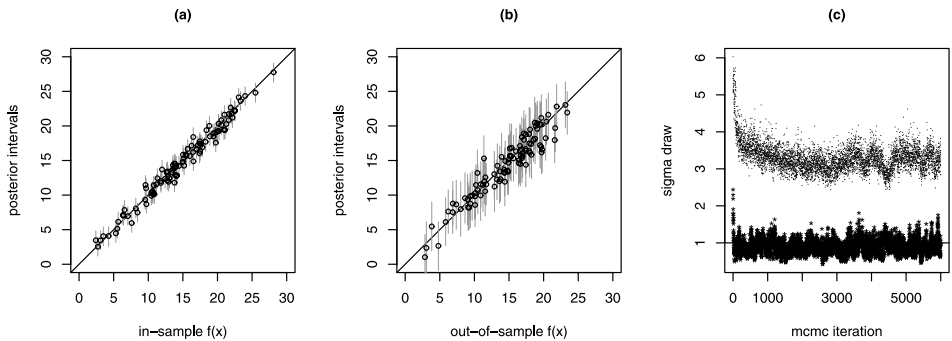}

 \caption{Inference about Friedman's $f(x)$ in $p=10$ dimensions.}\label{fig:friedman}
\end{figure}

To begin with, for each value of $x$, we obtained
posterior mean estimates $\hat{f}(x)$ of $f(x)$  by averaging the
5000 $f^*(x)$ values as in (\ref{eq:fhat}). Endpoints of 90\% posterior intervals for
each $f(x)$ were obtained as the 5\% and 95\% quantiles of the
$f^*$ values.  Figure~\ref{fig:friedman}(a)
plots $\hat{f}(x)$ against
$f(x)$ for the $n= 100$ in-sample values of $x$ from
(\ref{eq:fri-xs}) which were used to generate the $y$ values using
$(\ref{eq:fri-ys})$. Vertical lines indicate the 90\% posterior
intervals for the $f(x)$'s. Figure~\ref{fig:friedman}(b) is the
analogous plot at 100 randomly selected out-of-sample $x$ values.
We see that in-sample the $\hat{f}(x)$ values correlate very well with the
true $f(x)$ values and the intervals tend to cover the true
values.  Out-of sample, there is a slight degradation of the correlation and
wider intervals indicating
greater uncertainty about $f(x)$ at new $x$ values.

Although one would not expect the 90\% posterior intervals to exhibit
90\% frequentist coverage, it may be of interest to note that 89\%
and 96\% of the intervals in Figures~\ref{fig:friedman}(a) and (b)
covered the true $f(x)$ value, respectively.   In fact, in over
200 independent replicates of this example we found average
coverage rates of  87\% (in-sample) and 93\% (out-of-sample).
In real data settings where $f$ is unknown,
bootstrap and/or cross-validation methods might be helpful to get similar
calibrations of frequentist coverage. It should be
noted, however, that for extreme $x$ values, the prior may exert more shrinkage toward 0, leading
to lower coverage frequencies.

The lower sequence in Figure~\ref{fig:friedman}(c) is the sequence
of $\sigma$ draws over the entire 1000 burn-in plus 5000
iterations (plotted with *). The horizontal line is drawn at the
true value $\sigma = 1$. The Markov chain here appears to reach
equilibrium quickly, and although there is autocorrelation, the
draws of $\sigma$ nicely wander around the true value $\sigma = 1$,
suggesting that we have fit but not overfit.   To further
highlight the deficiencies of a single tree model, the upper
sequence (plotted with $\cdot$) in Figure~\ref{fig:friedman}(c) is
a sequence of $\sigma$ draws when $m =1$, a single tree model, is used.  The sequence
seems to take longer to reach equilibrium and remains
substantially above the true value $\sigma = 1$.  Evidently a
single tree is inadequate to fit this data.

\begin{figure}[b]

\includegraphics{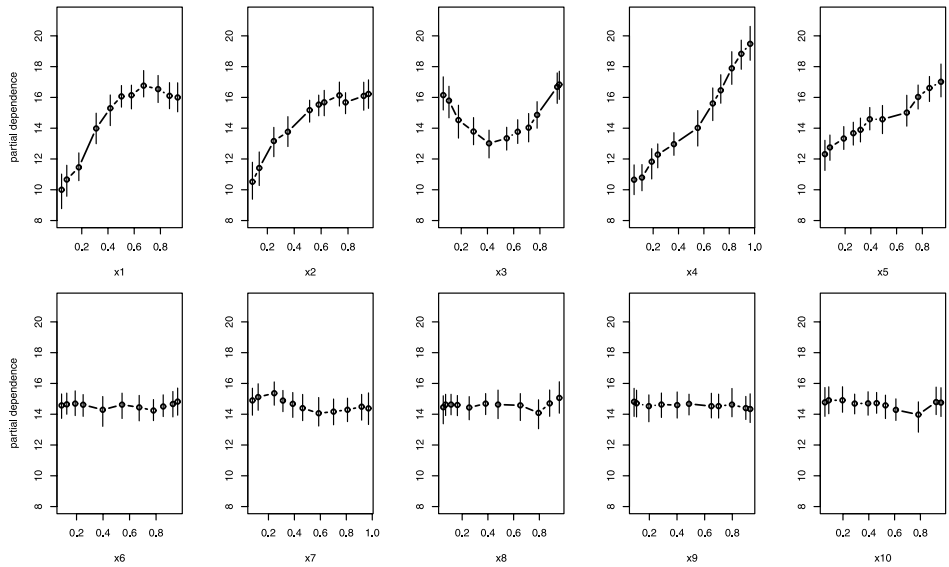}

\caption{Partial dependence plots for the 10 predictors in the Friedman data.}\label{fig:friedman-pdplot}
\end{figure}

Moving beyond estimation and inference about the values of $f(x)$, BART
estimates of the partial dependence functions $f(x_i)$ in
(\ref{eq:pdsum}) reveal the marginal effects of the individual $x_i$'s
on $y$. Figure \ref{fig:friedman-pdplot} shows the plots of point and
interval estimates of the partial dependence functions for
$x_1,\ldots,x_{10}$  from the 5000 MCMC samples of $f^*$.  The nonzero
marginal effects of $x_1,\ldots,x_5$ and the zero marginal effects of
$x_6,\ldots,x_{10}$ seem to be completely consistent with the form of
$f$ which of course would be unknown in practice.

\begin{figure}

\includegraphics{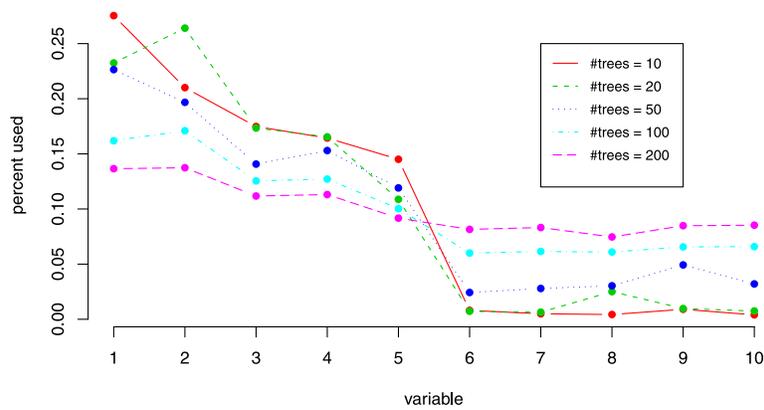}

\caption{Average use per splitting rule for variables $x_1,\ldots,x_{10}$ when $m =
10$, 20, 50, 100, 200.} \label{fig:friedman-varsel}
\end{figure}

As described in Section \ref{sec:estpd}, BART can also be used to
screen for variable selection by identifying as promising those
variables that are used most frequently in the sum-of-trees model $f^*$
draws from the posterior.   To illustrate the potential of this
approach here, we recorded the average use measure $v_i$ in
(\ref{eq:xfreq}) for each $x_i$ over 5000 MCMC draws of $f^*$ for each
of various values of $m$, based on a sample of $n = 500$ simulated
observations of the Friedman function (\ref{eq:fri-xs}) and
(\ref{eq:fri-ys}) with $p = 10$.  Figure \ref{fig:friedman-varsel}
plots these $v_i$ values for $x_1,\ldots,x_{10}$ for $m =
10$, 20, 50, 100, 200.  Quite dramatically, as the number of trees $m$ is
made smaller, the fitted sum-of-trees models increasingly incorporate
only those $x$ variables, namely, $x_1,\ldots,x_5$, that are needed to
explain the variation of $y$.  Without making use of any assumptions or
information about the actual functional form of $f$ in
(\ref{eq:fri-ys}), BART has here exactly identified the subset of
variables on which $f$ depends.

\begin{figure}

\includegraphics{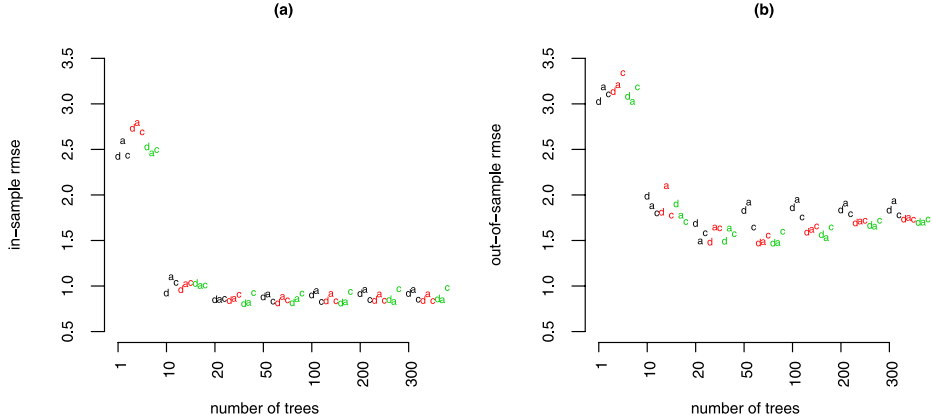}

 \caption{BART's robust RMSE performance as $(\nu, q, k, m)$ is varied
[a/d/c correspond to aggressive/default/conservative prior on $\sigma$,
black/red/green correspond to $k=(1,2,3)$]: \textup{(a)} in-sample RMSE
comparisons and \textup{(b)} out-of-sample RMSE comparisons. (Horizontal
jittering of points has been used to improve readability).}\label{fig:friedmanmrun1}
\end{figure}

Yet another appealing feature of BART is its apparent
robustness to small changes in the prior and to the
choice of $m$, the number of trees.   This robustness is illustrated in
Figures~\ref{fig:friedmanmrun1}(a) and (b) which display the in-
and out-of-sample RMSE  obtained by BART over 5000 MCMC samples of $f^*$ for various choices of $(\nu,
q, k, m )$. In each plot of RMSE versus $m$, the plotted text
indicates the values of $(\nu, q, k)$: $k = 1$, 2 or 3 and $(\nu,
q) = \mathrm{d}$, a or c (default/agressive/conservative). Three striking
features of the plot are apparent: (i) a very small number of
trees ($m$ very small) gives poor RMSE results, (ii) as long as $k>1$,
very similar results are obtained from different prior settings,
and (iii) increasing the number of trees well beyond the number
needed to capture the fit results in only a slight degradation of
the performance.

As Figure~\ref{fig:friedmanmrun1} suggests, the BART fitted values
are remarkably stable as the settings are varied. Indeed, in this
example, the correlations between out-of-sample fits turn out to
be very high, almost always greater than 0.99. For example, the
correlation between the fits from the $(\nu,q,k,m)=(3,0.9,2,100)$
setting (a reasonable default choice) and the $(10,0.75,3,100)$
setting (a very conservative choice) is 0.9948. Replicate runs with
different seeds are also stable: The correlation between fits from
two runs with the $(3,0.9,2,200)$ setting is 0.9994. Such stability
enables the use of one long MCMC run.  In contrast, some models
such as neural networks require multiple starts to ensure a good
optimum has been found.

\subsubsection{Finding low dimensional structure in high dimensional data}\label{sec:friedman-finding}

Of the $p$ variables $x_1,\ldots,x_p$ from (\ref{eq:fri-xs}), $f$
in (\ref{eq:fri-ys}) is a function of only five $x_1,\ldots,x_5$.
Thus, the problem we have been considering is one of drawing
inference about a five dimensional signal embedded in a $p$
dimensional space. In the previous subsection we saw that when $p
= 10$, the setup used by \citeasnoun{Frie1991}, BART could easily
detect and draw inference about this five dimensional signal with
just $n = 100$ observations. We now consider the same problem with
substantially larger values of $p$ to illustrate the extent to
which BART can find low dimensional structure in high dimensional
data.   For this purpose, we repeated the analysis displayed in
Figure~\ref{fig:friedman} with $p = 20$, 100 and 1000 but again
with only $n = 100$ observations.  We used BART with the same
default setting of $(\nu, q, k) = (3,0.90,2)$ and $m = 100$ with
one exception: we used the naive estimate $\hat{\sigma}$ (the
sample standard deviation of $Y$) rather the least squares
estimate to anchor the $q$th prior quantile to allow for data with
$p \ge n$. Note that because the naive $\hat{\sigma}$ is very
likely to be larger than the least squares estimate, it would also
have been reasonable to use a more aggressive prior setting for
$(\nu,q)$.

\begin{figure}

\includegraphics{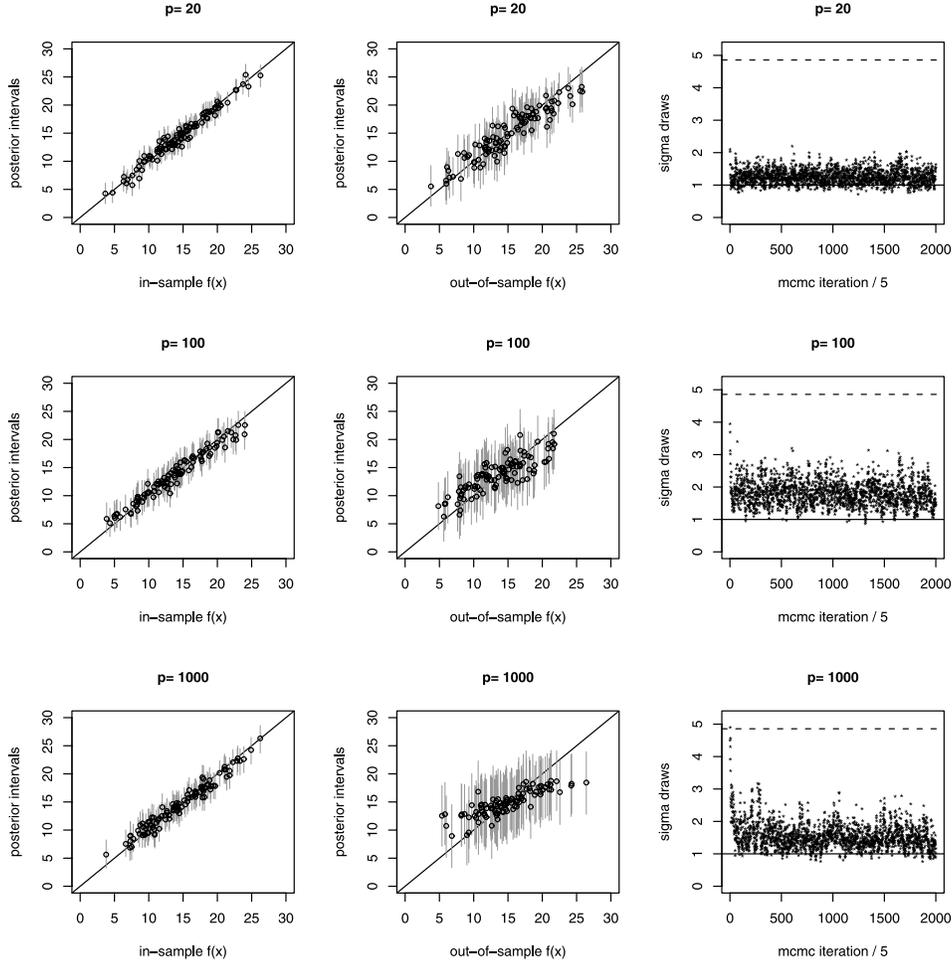}

\caption{Inference about Friedman's function in $p = 20$, 100, 1000 dimensions.}\label{fig:friedmanbigp}
\end{figure}

Figure~\ref{fig:friedmanbigp} displays the in-sample and
out-of-sample BART inferences for the larger values $p = 20$, 100
and 1000. The in-sample estimates and 90\% posterior intervals for
$f(x)$ are remarkably good for every $p$.  As would be expected,
the out-of-sample plots show that extrapolation outside the data
becomes less reliable as $p$ increases.  Indeed, the estimates
are shrunk toward the mean more, especially when $f(x)$ is near an extreme, and the
posterior intervals widen (as they should). Where there is less
information, it makes sense that BART pulls toward the center
because the prior takes over and the $\mu$'s are shrunk toward
the center of the $y$ values.  Nonetheless, when the dimension
$p$ is so large compared to the sample size $n = 100$, it is
remarkable that the BART inferences are at all reliable,
at least in the middle of the data.

In the third column of Figure~\ref{fig:friedmanbigp}, it is
interesting to note what happens to the MCMC sequence of $\sigma$
draws. In each of these plots, the solid line at $\sigma = 1$ is
the true value and the dashed line at $\hat\sigma = 4.87$ is the
naive estimate used to anchor the prior.  In each case, the
$\sigma$ sequence repeatedly crosses $\sigma = 1$. However, as $p$
gets larger, it increasingly tends to stray back toward larger
values, a reflection of increasing uncertainty. Last, note that
the sequence of $\sigma$ draws in Figure~\ref{fig:friedmanbigp}
is systematically higher than the $\sigma$ draws in
Figure~\ref{fig:friedman}(c). This may be due in part to the fact that
the regression $\hat\sigma$ rather than the naive $\hat\sigma$ was
used to anchor the prior in Figure~\ref{fig:friedman}. Indeed, if
the naive $\hat\sigma$ was instead used for
Figure~\ref{fig:friedman}, the $\sigma$ draws would similarly
rise.

A further attractive feature of BART is that it appears to avoid
being misled by pure noise.  To gauge this, we simulated $n = 100$
observations from (\ref{eq:fri-xs}) with $f \equiv 0$ for
$p=10$, 100, 1000 and ran BART with the same settings as above. With
$p=10$ and $p=100$ all intervals for $f$ at both in-sample and
out-of-sample $x$ values covered or were close to 0, clearly
indicating the absence of a relationship. At $p=1000$ the data
becomes so uninformative that our prior, which suggests that there
is some fit, takes over and some in-sample intervals are far from
0. However, the out-of-sample intervals still tend to cover 0 and
are very large so that BART still indicates  no evidence of a
relationship between $y$ and $x$.

\subsubsection{Out-of-sample comparisons with competing methods}\label{sec:friedman-train-test}

To gauge how well BART performs on the Friedman setup,
we compared its out-of-sample performance with random forests, neural nets and gradient boosting.
We dropped the Lasso since it has no hope of uncovering the nonlinear structure without
substantial modification of
the approach we used in Section~\ref{sec:bakeoff}.
In the spirit of Section~\ref{sec:friedman-finding}, we consider the case of estimating $f$
with just $n = 100$ observations when $p = 10$, 100 and 1000.
For this experiment we based both the BART-default and BART-cv estimates on 3000 MCMC
iterations obtained after 1000 burn-in draws.

For each value of $p$, we simulated 100 data sets of $n = 100$ observations each.
As in Section~\ref{sec:bakeoff}, we used 5-fold cross-validation to choose tuning parameters.
Because $f$ is known here, there was no need to simulate test set data.
Rather, for each method's $\hat{f}$ based on each data set, we randomly drew 1000 independent $x$ values
and assessed the fit using
$\mathrm{RMSE} = \sqrt{\frac{1}{1000} \sum_{i=1}^{1000} (\hat{f}(x_i) - f(x_i))^2}$.
For each method we thus obtained 100 such RMSE values.

For $p=10$, we used the same parameter values given in
Table~\ref{tab:parameters} for all the methods. For $p = 100$ and
1000, as in Section \ref{sec:friedman-finding}, we based the BART
prior for $\sigma$ on the sample standard deviation of $y$ rather than
on the least squares estimate.  For $p=100$, we changed the settings
for neural nets.  We considered either 3 or 6 hidden units and decay
values of 0.1, 1, 2, 3, 5, 10 or 20.  With the larger value of $p$, neural
nets use far more parameters so we had to limit the number of units and
increase the shrinkage in order to avoid consistently hitting a
boundary. At $p = 1000$, computational difficulties forced us to drop
neural nets altogether.

\begin{figure}

\includegraphics{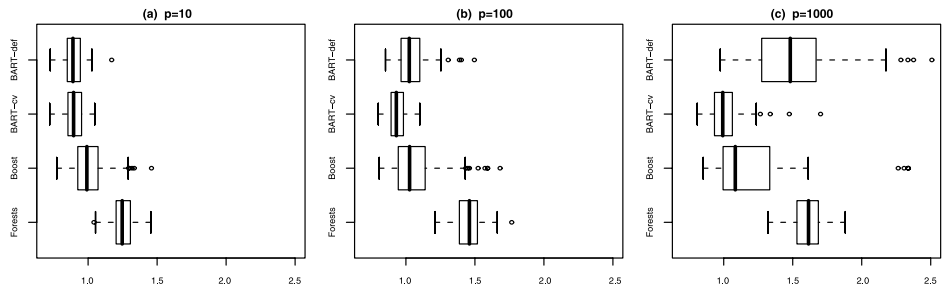}

\caption{Out-of-sample predictive comparisons in the Friedman simulated example for
(from top to bottom)  BART-default, BART-cv, boosting and random forests.
Each boxplot represents 100 RMSE values.} \label{fig:friedman-out}
\end{figure}

Figure~\ref{fig:friedman-out} displays boxplots and
Table~\ref{tab:friedman-perf} provides the 50\% and 75\% quantiles of
the 100 RMSE values for each method for $p = 10$, 100 and 1000.  (Note
that these are not relative RRMSE values as we had used in Figure
\ref{fig:boxplot}.)  With $p = 10$, the two BART approaches are clearly
the best and very similar.  However, as $p$ increases, BART-cv degrades
relatively little, whereas BART-default gets much worse. Indeed, when
$p = 1000$, BART-cv is much better than the other methods and the
performance of BART-default is relatively poor.

Evidently, the default prior is not a good choice for the Friedman simulation when $p$ is large.
This can be seen by noting that
in the cross-validation selection of tuning parameters for BART-cv,
the setting with $m = 50$ trees and the aggressive prior on $\sigma$ ($\mathrm{df}=3$, $\mathrm{quantile}=0.99$) is
chosen 60\% of the time when $p = 100$ or 1000.  Because of a high signal-to-noise ratio here,
the default $\sigma$ prior settings are apparently not aggressive enough
when the sample standard deviation of $y$ is used to anchor the quantile.  Furthermore,
since only five of the variables actually matter, $m = 50$ trees is adequate to fit
the complexity of the true $f$, whereas using more trees may inhibit the stochastic
search in this very high dimensional problem.

\begin{table}[b]
\tablewidth=290pt
\caption{(50\%, 75\%) quantiles of RMSE values for each method when $p = 10$, 100, 1000}\label{tab:friedman-perf}
\begin{tabular*}{290pt}{@{\extracolsep{\fill}}lccc@{}}
\hline
\textbf{Method}         &   $\bolds{p = 10}$        & $\bolds{p = 100}$        & $\bolds{p = 1000}$       \\
\hline
Random forests &  (1.25, 1.31) & (1.46, 1.52) & (1.62, 1.68)  \\
Neural net     &  (1.01, 1.32) & (1.71, 2.11) &   unavailable  \\
Boosting       &  (0.99, 1.07) & (1.03, 1.14) & (1.08, 1.33)  \\
BART-cv        &  (0.90, 0.95) & (0.93, 0.98) & (0.99, 1.06)  \\
BART-default   &  (0.89, 0.94) & (1.02, 1.10) & (1.48, 1.66)\\
\hline
\end{tabular*}
\end{table}

\subsection{Classification: A drug discovery application}\label{sec:drugdisc}

Our last example illustrates an application of the BART probit approach
of Section \ref{sec:classification} to a drug discovery
classification problem.  In such problems, the goal is to predict
the ``activity'' of a compound using
predictor variables that characterize the molecular structure of the
compound.  By ``activity,'' one typically means the ability to effect
a desired outcome against some biological target, such as inhibiting or
killing a certain virus.

The data we consider describe $p = 266$ molecular characteristics
of $n = 29\mbox{,}374$ compounds, of which 542 were classified as active.
These predictors represent topological aspects of molecular structure.
This data set was collected by the National Cancer Institute, and is
described in \citeasnoun{Feng2003}.  Designating the activity of a compound
by a binary variable ($Y=1$ if active and $Y=0$ otherwise), BART probit can be applied here
to obtain posterior mean estimates of  $P[Y = 1  | x]$ for each $x$ vector of the 266
molecular predictor values.

To get a feel for the extent to which BART's  $P[Y = 1  | x]$ estimates
can be used to identify promising drugs, we randomly split the data
into nonoverlapping train and test sets, each with 14,687 compounds of
which 271 were active.  We then applied BART probit to the training set
with the default settings $m = 50$ trees and mean shrinkage $k = 2$
(recall $\nu$ and $q$ have no meaning for the probit model).  To gauge
MCMC convergence, we performed four independent repetitions of 250,000
MCMC iterations and obtained essentially the same results each time.

\begin{figure}

\includegraphics{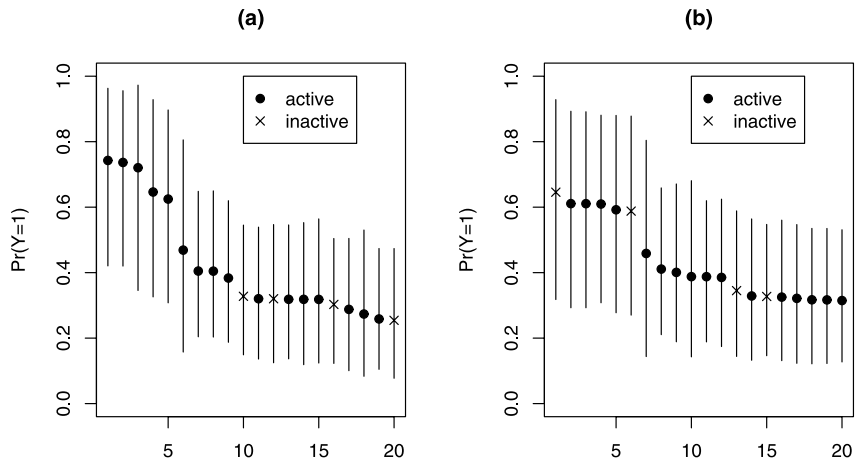}

\caption{BART posterior intervals for the 20 compounds with highest
predicted activity, using train~\textup{(a)} and test \textup{(b)} sets.}\label{intervals}
\end{figure}

Figure~\ref{intervals} plots the 20 largest  $P[Y = 1  | x]$ estimates
for the train and the test sets.  Also provided are the 90\% posterior
intervals which convey uncertainty and the identification whether the
drug was in fact active ($y = 1$) or not ($y = 0$).  The true positive
rates in both the train and test sets for these 20 largest estimates
are $16/20 = 80$\% (there are 4 inactives in each plot), an impressive
gain over the $271/14\mbox{,}687 = 1.85$\% base rate.  It may be of interest to
note that the test set intervals are slightly wider, with an average
width of 0.50 compared to 0.47 for the training intervals.

To gauge the predictive performance of BART probit on this data, we
compared its out-of sample performance with  boosted trees, neural
networks and random forests (using \texttt{gbm}, \texttt{nnet} and
\texttt{randomforest}, as in Section \ref{sec:bakeoff}) and with
support vector machines [using \texttt{svm} in the \texttt{e1071}
package of \citeasnoun{e1071}].  L1-penalized logistic regression was
excluded due to numeric difficulties.   For this purpose, we randomly
split the data into training and test sets, each containing 271
randomly selected active compounds.  The remaining inactive compounds
were then randomly allocated to create a training set of 1000 compounds
and a test set of 28,374 observations.  The training set was
deliberately chosen smaller to make feasible a comparative experiment
with 20 replications.

For this experiment we considered both BART-default and BART-cv based
on 10,000 MCMC iterations.  For BART-default, we used the same default
settings as above, namely, $m=200$ trees and $k=2$.  For BART-cv, we
used 5-fold cross-validation to choose from among $k=0.25$, 0.5, 1, 2,
3 and $m = 100$, 200, 400 or 800.  For all the competitors, we also
used 5-fold cross-validation to select tuning parameters as in Section
\ref{sec:bakeoff}.  However, the large number of predictors led to some
different ranges of tuning parameters.   Neural networks utilized a
skip layer and 0, 1 or 2 hidden units, with possible decay values of
0.0001, 0.1, 0.5, 1, 2, 5, 10, 20 and 50.  Even with 2 hidden units,
the neural network model has over 800 weights.  In random forests, we
considered 2\% variable sampling in addition to 10\%, 25\%, 50\% and
100\%. For support vector machines, two parameters, $C$, the cost of a
constraint violation, and $\gamma$ [\citet{CC01a}], were chosen by
cross-validation, with possible values $C=2^a, a=-6, -5, \ldots, 0$ and
$\gamma = 2^b, b=-7, -6, -5, -4$.

In each of 20 replicates, a different train/test split was generated.
Test set performance for this classification problem was measured by area under the
Receiver Operating Characteristic (ROC) curve, via the ROCR package of
\citeasnoun{ROCR}.  To generate a ROC curve, each method must produce a rank
ordering of cases by predicted activity.  All models considered generate a
predicted probability of activity, though other rank orderings could be
used.  Larger AUC values indicate superior performance, with an AUC of 0.50
corresponding to the expected performance of a method that randomly
orders observations by their predictions.  A classifier's AUC value
is the probability that it will rank a randomly chosen $y = 1$ example
higher than a randomly chosen $y= 0$. 

\begin{table}[b]
\caption{Classifier performance for the drug discovery problem, measured as
AUC, the area under a ROC curve.  Results are averages over 20 replicates.
The corresponding standard error is 0.0040, based~on an ANOVA of
AUC scores with a block effect for replicates}\label{tab:AUC}
\begin{tabular*}{6cm}{@{\extracolsep{\fill}}lc@{}}
\hline
\textbf{Method} & \textbf{AUC} \\
\hline
Random forests & 0.7680\\
Boosting       & 0.7543\\
BART-cv        & 0.7483\\
Support vector & 0.7417\\
BART           & 0.7245\\
Neural network & 0.7205\\
\hline
\end{tabular*}
\end{table}

The area under curve (AUC) values in Table~\ref{tab:AUC} indicate that for
this data set, BART is very competitive with all the methods.  Here
random forests provides the best performance, followed closely by
boosting, BART-cv and then support vector machines.  The default version of BART
and neural networks score slightly lower.  Although the differences in AUC between
these three groups are statistically significant (based on a 1-way ANOVA with a
block effect for each replicate), the practical differences are not appreciable.
We remark again that by avoiding the cross-validated selection of tuning
parameters, BART-default is much faster and easier to implement than the other methods here.

Finally, we turn to the issue of variable selection and demonstrate
that by decreasing the number of trees $m$, BART probit can be used,
just as BART in Section~\ref{sec:friedman-simple}, to identify those predictors which have
the most influence on the \mbox{response}.  For this purpose, we modify the
data setup as follows: instead of holding out a test set, all 542 active
compounds and a subsample of 542 inactives were used to build a model.
Four independent chains, each with 1,000,000 iterations, were used.  The
large number of iterations was used to ensure stability in the
``percent usage'' variable selection index (\ref{eq:xfreq}).
BART probit with $k=2$ and
with $m = 5, 10, 20$ trees were considered.

\begin{figure}
\begin{tabular}{@{}cc@{}}

\includegraphics{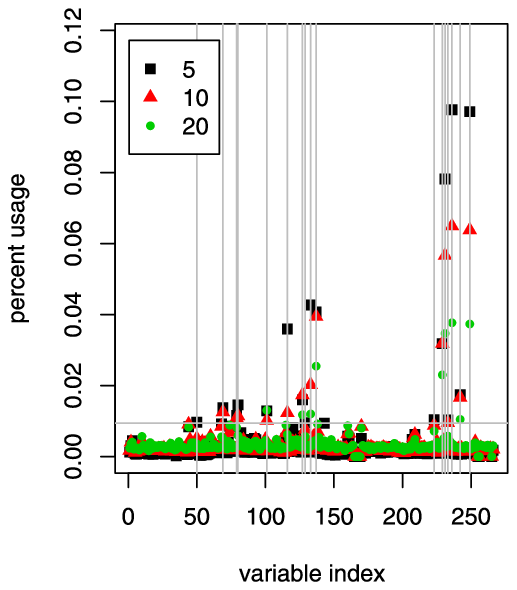}
&\includegraphics{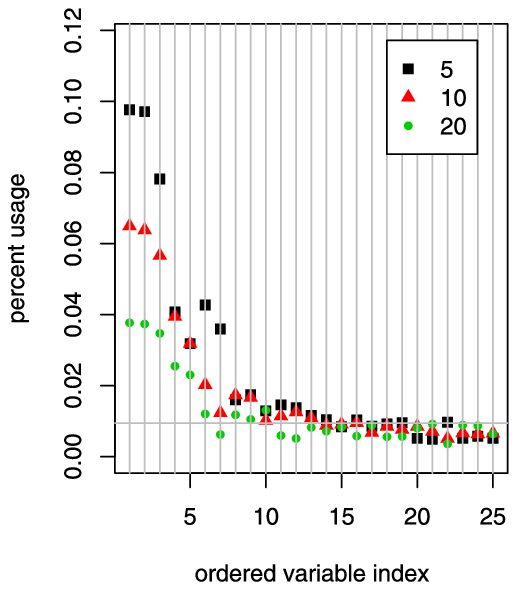}\\
(a)&(b)
\end{tabular}
 \caption{Variable importance measure, drug discovery example.  Values are
given for 5, 10 and 20 trees in the ensemble, for all 266 variables \textup{(a)} and
the 25 variables with the highest mean usage~\textup{(b)}.  Vertical lines in \textup{(a)}
indicate variables whose percent usage exceeds the 95th percentile.  The
95th percentile is indicated by a horizontal line.}\label{fig:usage}
\end{figure}

As Figure~\ref{fig:usage} shows,
the same three variables are selected as most
important for all three choices of $m$.
Considering that $1/266 \approx 0.004$, percent usages of 0.050 to
0.100 are quite a bit larger than one would expect if all variables were
equally important.  As expected, variable usage is most concentrated in the
case of a small ensemble (i.e., $m=5$ trees).

\section{Execution time considerations}\label{sec:executiontime}

In this section we study BART's execution time on various simulations of the Friedman data
 in order to shed light on how it depends on the sample size $n$ and
number of predictors $p$, and on how it compares to the execution time
of random forests, gradient boosting and neural nets.

\begin{figure}

\includegraphics{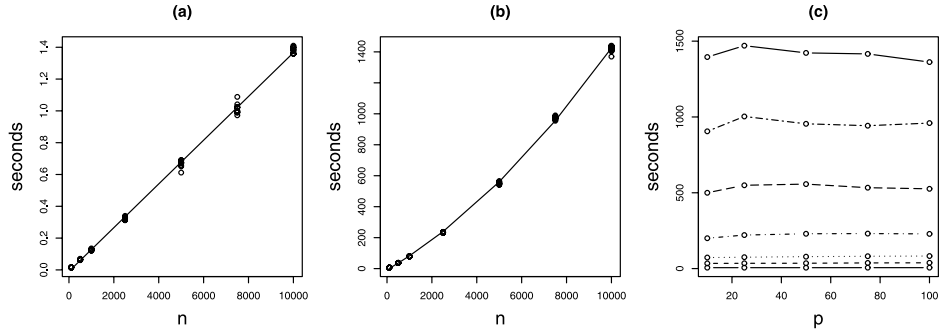}

\caption{\textup{(a)} For $p = 50$, execution times of the short
version of BART for $n= 100$, 500, 1000, 2500, 5000, 7500, 10,000, with a linear
regression overlaid. \textup{(b)} For $p = 50$, execution times
of the default version of BART for $n= 100$, 500, 1000, 2500, 5000, 7500, 10,000,  with a quadratic
regression overlaid.
\textup{(c)} Execution times for the default version of BART
when $p = 10$, 25, 50, 75, 100 for each $n= 100$, 500, 1000, 2500, 5000, 7500, 10,000.}\label{fig:times}
\end{figure}

To study the dependence of execution time on sample size $n$, we fixed
$p = 50$ and varied $n$ from 100 to 10,000.  For each $n$, we ran both
a short version (no burn-in iterations, 2 sampling iterations, $m =
200$ trees) and the default version (100 burn-in iterations, 1000
sampling iterations, $m = 200$ trees) of BART 10 times. The execution
times of these 10 replicates for each $n$  are displayed in
Figures~\ref{fig:times}(a) and (b). (We used the R \texttt{system.time}
command to time each run). Replicate variation is negligible. Because
BART's main computational task is the calculation of residuals in
(\ref{newdraw}) and the evaluation of log-likelihood in the
Metropolis--Hastings proposal, both of which involve iterating over
either all $n$ observations or all observations contained in a node, we
anticipated that execution time would increase linearly with $n$.  This
linearity was indeed borne out by the short version of BART in
Figure~\ref{fig:times}(a).

However, for the longer default version of BART, this dependence
becomes quadratic as is evidenced in Figure~\ref{fig:times}(b).
Apparently, this nonlinear dependence is due to the adaptive nature of
BART.  For larger $n$, BART iterations tend toward the use of larger
trees to exploit finer structure, and these larger trees require more
tree-based operations to generate the predictions required for residual
and likelihood evaluation.  Indeed, in a separate experiment using $m =
50$ trees, we found that for $n = 100$, BART trees had up to 4 terminal
nodes with an average size of 2.52 terminal nodes, whereas for $n =
10\mbox{,}000$, BART trees had as many as 10 terminal nodes with an average
size of 3.34. In contrast, the short version BART effectively keeps
tree sizes small by limiting iterations, so that its execution time
scales linearly with $n$.

To study the dependence of execution time on the number of predictors
$p$, we replicated the above experiment for the default version of BART
varying $p$ from 10 to 100 for each $n$.  The execution times,
displayed in Figure~\ref{fig:times}(c), reveal that in all cases,
BART's execution time is close to independent of $p$, especially as
compared to its dependence on $n$.  Note, however, that, in practice,
the time to run BART may depend on the complexity of the underlying
signal which may require a longer burn-in period  and a longer set of
runs to fully explore the posterior.  Larger values of $p$ may lead to
such complexity.

Finally, we compared BART's execution time to that of random forests,
gradient boosting and neural nets, where execution of each method
entails generating predictions for the training set.  As in our first
experiment above, we fixed $p=50$ and varied $n$ from 100 to 10,000.
Two versions of BART were run: the default version considered above and
a minimal version (20 burn-in iterations, 10 sampling iterations, $m =
50$ trees).  Even with such a small number of iterations, the fits
provided by this minimal version were virtually indistinguishable from
the default version for the Friedman data with $n=100$ and $p=10$. For
the other models, tuning parameters were held fixed at the ``typical''
values: \texttt{mtry = 10} and \texttt{ntree = 500} for
\texttt{RandomForest}; \texttt{shrinkage = 0.1},
\texttt{interaction.depth = 3} and \texttt{n.tree = 100} for
\texttt{gbm};  \texttt{size = 6} and  \texttt{decay = 1.0} for
\texttt{nnet}.

\begin{figure}[b]

\includegraphics{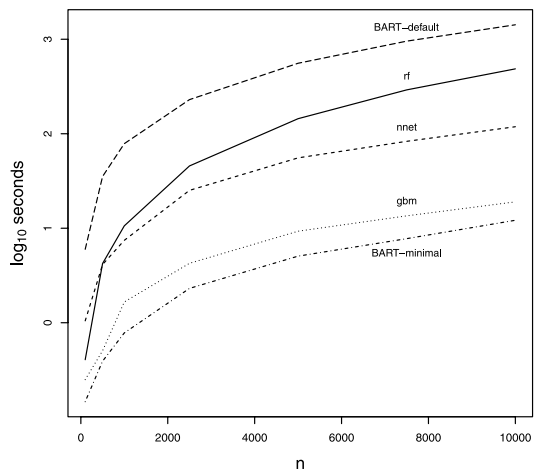}

\caption{Execution time comparisons of various methods, with $\log_{10}$ seconds plotted versus sample
size $n= 100$, 500, 1000, 2500, 5000, 7500, 10,000.}
\label{fig:compare}
\end{figure}

Execution times as a function of $n$ for each of the methods are
displayed in Figure~\ref{fig:compare}.   The execution time of BART is
seen to be comparable with that of the other algorithms, and all the
algorithms scale in a similar fashion.  The minimal version of BART is
faster than all the other algorithms, while the default version is the
slowest.  Of course,  execution times under actual use should take into
account the need to select tuning parameters, typically by
cross-validation.   By being competitive while avoiding this need, as
was illustrated in Section~\ref{sec:bakeoff}, the default version of
BART compares most favorably with these other methods.

\section{Extensions and related work}\label{sec:related}

Although we have framed BART as a stand alone procedure, it can
also be incorporated into larger statistical models, for example,
by adding other components such as linear terms or linear random
effects.  For instance,
one might consider a model of the form
\begin{equation}\label{h2}
Y = h_1(x)  + h_2(z) +  \epsilon, \qquad  \epsilon\sim N(0,\sigma^2),
\end{equation}
where $h_1(x)$ is a sum of trees as in (\ref{sstmodel1}) and $h_2(z)$ is a parametric form involving~$z$, a second vector of predictors.
One can also extend the sum-of-trees model to a
multivariate framework such as
\begin{equation}\label{multivarBART}
Y_i = h_i(x_i) + \epsilon_i,\qquad
(\epsilon_1,\epsilon_2,\ldots,\epsilon_p) \sim N(0,\Sigma),
\end{equation}
where each $h_i$ is a sum of trees and $\Sigma$ is a $p$ dimensional
covariance matrix.  If all the $x_i$ are the same, we have a
generalization of multivariate regression. If the $x_i$ are
different, we have a generalization of Zellner's SUR model
[\citet{Zell1962}].  The modularity of the BART MCMC algorithm in
Section \ref{sec:mcmc} easily allows for such incorporations and
extensions.  Implementation of linear terms or random effects in a
BART model would only require a simple additional MCMC step to draw
the associated parameters. The multivariate version of BART
(\ref{multivarBART}) is easily fit by drawing each $h^*_i$ given
$\{h^*_j\}_{j \ne i}$ and $\Sigma$, and then drawing $\Sigma$ given
all the $h^*_i$.

The framework for variable selection developed in Section
\ref{sec:postcalc} and illustrated in Section \ref{sec:examples}
appears quite promising for model-free identification of important
features.   Modification of the prior hyperparameters may further
enhance this approach.  For instance, in the tree prior
(\ref{treeprior}), the default $\alpha=0.95$ puts only 5\% prior
probability on a single node tree.  This may encourage splits even in
situations where predictive gains are modest.  Putting more mass on
small trees (via smaller values of $\alpha$) might lead to a posterior
in which ``every split counts,'' offsetting the tendency of BART to
include spurious splits.  Although such spurious splits do not affect
predictive accuracy, they do tend to inflate variable usage
frequencies, thereby making it more difficult to distinguish the
important variables.  Prior specifcations for variable selection via
BART are part of our ongoing research.

An early version of our work on BART [\citet{ChipGeorMcCu2007}] was
published in the proceedings of the conference Advances in Neural
Information Processing Systems 2006.  Based on this and other
preliminary technical reports of ours, a variety of extensions and
applications of BART have begun to appear.  \citeasnoun{ZhaShiMul2007}
proposed SBART an extension of BART obtained by adding a spatial
component along the lines of (\ref{h2}).  Applied to the problem of
merging data sets, they found that SBART improved over the conventional
census based method.  For the predictive modeling problem of TF-DNA
binding in genetics, 
\citeasnoun{ZhoLiu2008} considered a variety of learning methods,
including stepwise linear regression, MARS, neural networks, support
vector machines, boosting and BART.  Concluding that ``the BART method
performed best in all cases,'' they noted BART's ``high predictive
power, its explicit quantification of uncertainty and its
interpretability.''  By keeping track of the per sample inclusion
rates, they successfully used BART to identify some unusual predictors.
\citeasnoun{ZhaHae2007} independently developed a probit extension
of BART, which they call BACT,  and applied it to credit risk data to
predict the insolvency of firms.  They found BACT to outperform the
logit model, CART and support vector machines.
\citeasnoun{AbuNapWan2008} also independently discovered the probit
extension of BART, which they call CBART, and applied it for the
automatic detection of phishing emails.  They found CBART to outperform
logistic regression, random forests, support vector machines, CART,
neural networks and the original BART. \citeasnoun{AbreMcCu2006}
applied BART to hockey game penalty data and found evidence of referee
bias in officiating. Without exception, these papers provide further
evidence for the remarkable potential of BART.

\section{Discussion}\label{sec:disc}

The essential components of BART are the sum-of-trees model, the
regularization prior  and the backfitting MCMC algorithm. As opposed to the
Bayesian approaches of CGM98 and \citeasnoun{DeniMallSmit1998},
where a single tree is used to explain all the variation in $y$,
each of the trees in BART accounts for only part of the overall fit.  This is accomplished with
a regularization prior that shrinks the tree effects toward a simpler fit.
To facilitate the implementation of BART, the prior is formulated in terms of
rapidly computable forms that are controlled by interpretable hyperparameters,
and which allow for a highly effective default version for
immediate ``off-the-shelf'' use.  Posterior calculation is carried out by a
tailored backfitting MCMC algorithm that appears to converge quickly, effectively
obtaining a (dependent) sample from the posterior distribution over the space of sum-of-trees models.
A variety of inferential quantities of interest can be obtained directly from this sample.

The application of BART to a wide variety of data sets and a simulation
experiment (Section \ref{sec:examples}) served to demonstrate many of
its appealing features. In terms of out-of sample predictive RMSE
performance, BART compared favorably with boosting, the lasso, MARS,
neural nets and random forests.  In particular, the computationally
inexpensive and easy to use default version of BART performed extremely
well. In the simulation experiments, BART obtained reliable posterior
mean and interval estimates of the true regression function as well as
the marginal predictor effects.  BART's performance was seen to be
remarkably robust to hyperparameter specification, and remained
effective when the regression function was buried in ever higher
dimensional spaces. BART was also seen to be a new effective tool for
model-free variable selection.  Finally, a straightforward probit
extension of BART for classification of binary $Y$ was seen to be an
effective, competitive tool for discovering promising drugs on the
basis of their molecular structure.

\printaddresses


\begin{thebibliography}{99}

\bibitem[\protect\citeauthoryear{Abreveya and McCulloch}{2006}]{AbreMcCu2006}
\textsc{Abreveya}, J. and \textsc{McCulloch}, R.  (2006).
  Reversal of fortune: A statistical analysis of penalty calls in the national
  hockey league. Technical report, Purdue Univ.

\bibitem[\protect\citeauthoryear{Abu-Nimeh et al.}{2008}]{AbuNapWan2008}
\textsc{Abu-Nimeh}, S., \textsc{Nappa}, D., \textsc{Wang}, X. and \textsc{Nair}, S.  (2008). Detecting phishing emails via {B}ayesian additive
  regression trees. Technical report, Southern Methodist Univ., Dallas,
  TX.

\bibitem[\protect\citeauthoryear{Albert and Chib}{1993}]{AlbeChib1993}
\textsc{Albert}, J. H. and \textsc{Chib}, S.  (1993). Bayesian
  analysis of binary and polychotomous response data. \textit{J.
  Amer. Statist. Assoc.} \textbf{88} 669--679.
\MR{1224394}

\bibitem[\protect\citeauthoryear{Amit and Geman}{1997}]{AmitGema1997}
\textsc{Amit}, Y. and \textsc{Geman}, D.  (1997). Shape
  quantization and recognition with randomized trees. \textit{Neural Computation}
  \textbf{9} 1545--1588.

\bibitem[\protect\citeauthoryear{Blanchard}{2004}]{Blan2004}
\textsc{Blanchard}, G.  (2004). Un algorithme accelere
  d'echantillonnage {B}ayesien pour le modele {CART}. \textit{Revue
  d'Intelligence artificielle} \textbf{18} 383--410.

\bibitem[\protect\citeauthoryear{Breiman}{1996}]{Br96a}
\textsc{Breiman}, L.  (1996). Bagging predictors. \textit{Machine Learning} \textbf{26} 123--140.

\bibitem[\protect\citeauthoryear{Breiman}{2001}]{Bre2001}
\textsc{Breiman}, L.  (2001). Random forests. \textit{Machine
  Learning} \textbf{45} 5--32.

\bibitem[\protect\citeauthoryear{Chang and Lin}{2001}]{CC01a}
\textsc{Chang}, C.-C. and \textsc{Lin}, C.-J.  (2001). LIBSVM: A library for support vector machines.
Available at \url{http://www.csie.ntu.edu.tw/~cjlin/libsvm}.

\bibitem[\protect\citeauthoryear{Chipman, George and
  McCulloch}{1998}]{ChipGeorMcCu1998a}
\textsc{Chipman}, H. A., \textsc{George}, E. I. and \textsc{McCulloch}, R. E.  (1998).
Bayesian {CART} model search (with discussion and a rejoinder by the authors). \textit{J. Amer. Statist. Assoc.} \textbf{93} 935--960.

\bibitem[\protect\citeauthoryear{Chipman, George and
  McCulloch}{2002}]{ChipGeorMcCu2002a}
\textsc{Chipman}, H. A., \textsc{George}, E. I. and \textsc{McCulloch}, R. E.  (2002).
Bayesian treed models. \textit{Machine Learning} \textbf{48} 299--320.

\bibitem[\protect\citeauthoryear{Chipman, George and
  McCulloch}{2007}]{ChipGeorMcCu2007}
\textsc{Chipman}, H. A., \textsc{George}, E. I. and \textsc{McCulloch}, R. E.  (2007). Bayesian ensemble learning. In \textit{Neural Information
  Processing Systems} \textbf{19} 265--272.

\bibitem[\protect\citeauthoryear{Denison, Mallick and
  Smith}{1998}]{DeniMallSmit1998}
\textsc{Denison}, D. G. T., \textsc{Mallick}, B. K. and \textsc{Smith}, A. F. M.  (1998).
A {B}ayesian {CART} algorithm. \textit{Biometrika}
  \textbf{85} 363--377.
\MR{1649118}

\bibitem[\protect\citeauthoryear{Dimitriadou et al.}{2008}]{e1071}
\textsc{Dimitriadou}, E., \textsc{Hornik}, K., \textsc{Leisch}, F., \textsc{Meyer}, D.  and \textsc{Weingessel},
  A.  (2008). e1071: Misc functions of the
  Department of Statistics (e1071), TU Wien.
\newblock R package version 1.5-18.

\bibitem[\protect\citeauthoryear{Efron et al.}{2004}]{EfrHasJohTib2004}
\textsc{Efron}, B., \textsc{Hastie}, T., \textsc{Johnstone}, I. and \textsc{Tibshirani}, R.  (2004).
Least angle regression (with discussion and a rejoinder by the authors). \textit{Ann. Statist.}
\textbf{32} 407--499.
\MR{2060166}

\bibitem[\protect\citeauthoryear{Feng et al.}{2003}]{Feng2003}
\textsc{Feng}, J., \textsc{Lurati}, L., \textsc{Ouyang}, H., \textsc{Robinson}, T., \textsc{Wang}, Y., \textsc{Yuan}, S. and
  \textsc{Young}, S.  (2003). Predictive toxicology:
  Benchmarking molecular descriptors and statistical methods. \textit{Journal of
  Chemical Information and Computer Sciences} \textbf{43} 1463--1470.

\bibitem[\protect\citeauthoryear{Freund and Schapire}{1997}]{FreSch1997}
\textsc{Freund}, Y. and \textsc{Schapire}, R. E.  (1997). A
  decision-theoretic generalization of on-line learning and an application to
  boosting. \textit{J. Comput. System Sci.} \textbf{55} 119--139.
\MR{1473055}

\bibitem[\protect\citeauthoryear{Friedman}{1991}]{Frie1991}
\textsc{Friedman}, J. H.  (1991). Multivariate adaptive
  regression splines (with discussion and a rejoinder by the author). \textit{Ann. Statist.} \textbf{19} 1--67.
\MR{1091842}

\bibitem[\protect\citeauthoryear{Friedman}{2001}]{Fri2001}
\textsc{Friedman}, J. H.  (2001). Greedy function
  approximation: A gradient boosting machine. \textit{Ann. Statist.}
  \textbf{29} 1189--1232.
\MR{1873328}

\bibitem[\protect\citeauthoryear{Green}{1995}]{Gre1995}
\textsc{Green}, P. J.  (1995). Reversible jump {MCMC}
  computation and {B}ayesian model determination. \textit{Biometrika} \textbf{82} 711--732.
\MR{1380810}

\bibitem[\protect\citeauthoryear{Hastie and Tibshirani}{2000}]{HastTibs2000}
\textsc{Hastie}, T. and \textsc{Tibshirani}, R.  (2000).
Bayesian backfitting (with comments and a rejoinder by the authors).
\textit{Statist. Sci.} \textbf{15} 196--223.
\MR{1820768}

\bibitem[\protect\citeauthoryear{Kim et al.}{2007}]{KimLohShiCha2007}
\textsc{Kim}, H., \textsc{Loh}, W.-Y., \textsc{Shih}, Y.-S. and \textsc{Chaudhuri}, P.  (2007). Visualizable and interpretable regression models with
  good prediction power. \textit{IEEE Transactions: Special Issue on Data Mining
  and Web Mining} \textbf{39} 565--579.


\bibitem[\protect\citeauthoryear{Ridgeway}{2004}]{Rid2004}
\textsc{Ridgeway}, G.  (2004). The gbm package. R
  Foundation for Statistical Computing, Vienna, Austria.

\bibitem[\protect\citeauthoryear{Sing et al.}{2007}]{ROCR}
\textsc{Sing}, T., \textsc{Sander}, O., \textsc{Beerenwinkel}, N. and \textsc{Lengauer}, T.  (2007).
ROCR: Visualizing the performance of scoring  classifiers.
R package version 1.0-2.

\bibitem[\protect\citeauthoryear{Venables and Ripley}{2002}]{VenaRipl2002}
\textsc{Venables}, W. N. and \textsc{Ripley}, B. D.  (2002).
\textit{Modern Applied  Statistics With {S}}, 4th ed. Springer, New York.

\bibitem[\protect\citeauthoryear{Wu, Tjelmeland and  West}{2007}]{WuTjeWes2007}
\textsc{Wu}, Y., \textsc{Tjelmeland}, H. and \textsc{West}, M.  (2007).
Bayesian {CART}: Prior specification and posterior simulation.
\textit{J. Comput. Graph. Statist.} \textbf{16} 44--66.
\MR{2345747}

\bibitem[\protect\citeauthoryear{Zellner}{1962}]{Zell1962}
\textsc{Zellner}, A.  (1962). An efficient method of
  estimating seemingly unrelated regressions and testing for aggregation
  bias. \textit{J. Amer. Statist. Assoc.} \textbf{57} 348--368.
\MR{0139235}

\bibitem[\protect\citeauthoryear{Zhang and Haerdle}{2010}]{ZhaHae2007}
\textsc{Zhang}, J. L. and \textsc{Haerdle}, W. K.  (2010). The
  {B}ayesian additive classification tree applied to credit risk
  modelling. \textit{Comput. Statist. Data Anal.} \textbf{54}
  1197--1205.

\bibitem[\protect\citeauthoryear{Zhang, Shih and
  Muller}{2007}]{ZhaShiMul2007}
\textsc{Zhang}, S., \textsc{Shih}, Y.-C. T. and \textsc{Muller}, P.  (2007). A spatially-adjusted {B}ayesian additive regression tree
  model to merge two datasets. \textit{Bayesian Anal.} \textbf{2} 611--634.
\MR{2342177}

\bibitem[\protect\citeauthoryear{Zhou and Liu}{2008}]{ZhoLiu2008}
\textsc{Zhou}, Q. and \textsc{Liu}, J. S.  (2008). Extracting
  sequence features to predict protein-{DNA} binding: A~comparative
  study. \textit{Nucleic Acids Research} \textbf{36} 4137--4148.

\end{thebibliography}
\end{document}